# MAX-PLUS DECOMPOSITION OF SUPERMARTINGALES AND CONVEX ORDER. APPLICATION TO AMERICAN OPTIONS AND PORTFOLIO INSURANCE

By Nicole El Karoui and Asma Meziou

*CMAP, Ecole Polytechnique*

We are concerned with a new type of supermartingale decomposition in the Max-Plus algebra, which essentially consists in expressing any supermartingale of class $(\mathcal{D})$ as a conditional expectation of some running supremum process. As an application, we show how the Max-Plus supermartingale decomposition allows, in particular, to solve the American optimal stopping problem without having to compute the option price. Some illustrative examples based on one-dimensional diffusion processes are then provided. Another interesting application concerns the portfolio insurance. Hence, based on the "Max-Plus martingale," we solve in the paper an optimization problem whose aim is to find the best martingale dominating a given floor process (on every intermediate date), w.r.t. the convex order on terminal values.

**1. Introduction.** One of the most important decompositions in modern probability theory is the Doob–Meyer decomposition of a supermartingale $Z$, into the unique difference of a local martingale $M$ and a predictable nondecreasing process $A$. This additive representation can also be written in the form of $Z_t = \mathbf{E}[A_\zeta - A_t|\mathcal{F}_t] + \mathbf{E}[Z_\zeta|\mathcal{F}_t]$, where $\zeta$ is a stopping time defining the horizon of the problem.

We provide in this paper another "additive" decomposition theorem for supermartingales of class $(\mathcal{D})$, in a nice mathematical structure called Max-Plus algebra. This latter is an idempotent semiring structure endowed with the two binary operations $\oplus = \max$ and $\otimes = +$. It introduces a linear algebra point of view to dynamic programming problems and large deviations, and turns out to be very effective to make algebraic computations (see [6]).

Specifically, given a quasi-left-continuous supermartingale $Z$ of class $(\mathcal{D})$, defined on $[0, \zeta]$ where $\zeta$ is a stopping time, our aim is to construct an optional upper-right semi-continuous process $L$ such that $Z$ can be represented in terms of the "running supremum" of $L$:

$$Z_t = \mathbf{E}\left[\sup_{t \leq u \leq \zeta} L_u \vee Z_\zeta | \mathcal{F}_t\right] = \mathbf{E}\left[\oint_{[t, \zeta]} L_u \oplus Z_\zeta | \mathcal{F}_t\right], \qquad 0 \leq t \leq \zeta.$$

Here $\oint$ denotes a nonlinear integral called Max-Plus integral. The running supremum of $L$ over $[0, \tau]$ for any stopping time $\tau$ between 0 and $\zeta$ can be seen as the value of a nondecreasing process $\Lambda_\tau$ and, hence, the above representation is analogous to the Doob–Meyer decomposition.

The martingale $M^\oplus$ defined as the conditional expectation of $\Lambda_\zeta \oplus Z_\zeta$ is called the martingale of the Max-Plus decomposition of $Z$ and dominates $\Lambda_\tau \oplus Z_\tau$ for any stopping time $\tau$. We first focus on the martingale uniqueness in Section 2 and show that the case is different for the nondecreasing process.

To establish the existence of such a decomposition, we essentially deal with some easy convex duality methods, which turned out to be particularly useful for the Bandit problem [19, 45], and for nonlinear representations of general processes [7, 15].

An interesting application of the Max-Plus decomposition of supermartingales is the problem of American Call options written on an underlying $Y$ of class $(\mathcal{D})$. In fact, based on the representation of the Snell envelope of $Y$ as a conditional expectation of some running supremum $L_{t,\zeta}^* = \sup_{t \leq u \leq \zeta} L_u$, we explicitly characterize an optimal stopping time in terms of the index process $L$ and, thus, the simple knowledge of $L$ completely solves the optimal stopping problem without having to compute the price of the American option. Moreover, thanks to the Max-Plus decomposition, we get an explicit characterization of American Call options as lookback ones. This particularly generalizes the results of Darling, Liggett and Taylor [10] on American options, only valid in the discrete case, where the underlying defines a partial sum of independent and identically distributed random variables with negative drift. Our extension to the case of geometric Lévy processes (with only negative jumps) leads to an index process $L$, proportional to the supermartingale underlying.

Our original interest in the Max-Plus decomposition of supermartingales comes from portfolio insurance which is a popular example of dynamic asset allocation. In fact, using the Max-Plus decomposition of supermartingales, the paper suggests a new approach to the classic utility maximization problem with American constraints. The optimization is performed with respect to stochastic convex ordering on the terminal value and this avoids arbitrary assumptions regarding the form of the utility function of a decision maker.

To this aim, we first consider a martingale constrained optimization problem, in terms of convex ordering. All admissible martingales must have the same initial value and must dominate a given floor process $Y$. Then we note that the martingale of the Max-Plus decomposition of the Snell envelope of $Y$ solves the addressed problem. We refer the interested reader to another paper [20] for more details about the financial application to portfolio insurance.

The paper is organized as follows. In the next section we extend the Doob–Meyer decomposition theorem for supermartingales in the Max-Plus algebra and use it to generalize in Section 3, the point of view of Darling and his co-authors concerning American optimal stopping problems. In Sections 3.2 and 3.3 we provide some illustrative examples based on multiplicative and additive Lévy processes, highlighting the link between the Max-Plus decomposition and American options. Then, we focus on a parameterized problem which allows us to derive an explicit decomposition for a given right-continuous supermartingale in the Max-Plus algebra. In Section 4.4 we apply our Max-Plus decomposition theorem to solve the American Call option problem, with a general underlying. In Section 4.5 we reconsider the different steps of the Max-Plus decomposition in a Markovian framework. Then in Section 5 we show that the martingale involved in this Max-Plus decomposition is also characterized as the optimal solution of a particular constrained optimization problem, expressed in terms of stochastic convex ordering. Finally, we give some closed formulae of martingales involved in the Max-Plus decomposition of multiplicative and additive Lévy processes. We then exploit the Azéma–Yor martingale (see [5]) to provide the Max-Plus decomposition of any concave increasing function of a continuous local martingale. This last example proves to be very insightful and covers, among others all the considered cases of Lévy processes. The more technical arguments are relegated to the Appendix.

## 2. Supermartingale decompositions.

2.1. *Framework.* We here give the notation and definitions which will be used throughout the paper. Uncertainty is modeled by some filtered probability space $(\Omega, \mathcal{F}, \mathbf{P})$ satisfying the usual conditions, that is, $\{\mathcal{F}_t\}$ is an increasing, right-continuous family of $\sigma$-fields and $\mathcal{F}_0$ contains all the $\mathbf{P}$-negligible events in $\mathcal{F}$. We further make the following assumption:

ASSUMPTION 2.1. $\{\mathcal{F}_t\}$ is assumed to be *quasi-left-continuous*, that is for any predictable stopping time $\tau$, one has $\mathcal{F}_\tau = \mathcal{F}_{\tau^-}$.

The *horizon* of the problem is a stopping time denoted by $\zeta$ and may be infinite.

An adapted process $X$ is said to be of class $(\mathcal{D})$ if $|X|$ is dominated by a uniformly integrable martingale. Another characterization of such processes, based on stopping times, is given in Section 4. Note that in the particular case where $X$ is a martingale, $X$ is of class $(\mathcal{D})$ if and only if it is uniformly integrable, u.i. in short. A supermartingale (martingale) $Z$ of class $(\mathcal{D})$ is called a $(\mathcal{D})$-supermartingale [$(\mathcal{D})$-martingale].

2.2. *Supermartingale decompositions.* One of the most fundamental decompositions in the theory of stochastic processes is the Doob–Meyer decomposition of a supermartingale as the unique difference between a local martingale and a predictable nondecreasing process. Recently, motivated by optimization problems in Mathematical Finance, such a decomposition has been extended to a process defining a supermartingale for an infinite family of equivalent probability measures. Nevertheless, El Karoui and Quenez [21], and Kramkov [34] have established that this decomposition only holds by relaxing the predictability property of the nondecreasing process. In this paper, for different reasons, we are also concerned with an optional supermartingale decomposition, but with a different meaning.

Moreover, the additive decomposition is not the only one to present an interest. For example, in Mathematical Finance, the multiplicative decomposition was paid particular attention [30]. We first recall the standard version of the Doob–Meyer decomposition theorem [31, 36] and [13], Theorem 4.10, page 24.

THEOREM 2.2. *Let $Z$ be a $(\mathcal{D})$-supermartingale defined on $[0, \zeta]$.*

1. *There exists a unique predictable càdlàg (right-continuous, left-limited) nondecreasing process $A$ such that $A_0 = 0$, $A_\zeta$ integrable, and*

$$(2.1) \qquad Z_t = \mathbf{E}[A_\zeta - A_t | \mathcal{F}_t] + \mathbf{E}[Z_\zeta | \mathcal{F}_t], \qquad 0 \le t \le \zeta.$$

2. *In addition, with this representation, we have*

$$(2.2) \qquad Z_t + A_t = M_t^A, \qquad 0 \le t \le \zeta,$$

*where $M^A$ defined by $M_t^A = \mathbf{E}[A_\zeta + Z_\zeta | \mathcal{F}_t]$ is a $(\mathcal{D})$-martingale. This decomposition is unique (up to indistinguishability).*

3. *Further, if $Z$ is quasi-left-continuous, then $A$ is continuous.*

As mentioned above, there exist other decompositions relative to operations that are different from addition, such as multiplication [29, 35].

THEOREM 2.3. *Let $Z$ be a positive $(\mathcal{D})$-supermartingale.*

1. *There exists a unique integrable predictable nondecreasing process $B$ satisfying $B_0 = 1$, such that*

$$(2.3) \qquad Z_t = \mathbf{E}\left[Z_\zeta \times \frac{B_\zeta}{B_t} \Big| \mathcal{F}_t\right], \qquad 0 \le t \le \zeta.$$

2. *In addition, with this representation, we have*

$$(2.4) \qquad Z_t \times B_t = M_t^P, \qquad 0 \le t \le \zeta,$$

*where $M^P$ defined by $M_t^P = \mathbf{E}[B_\zeta \times Z_\zeta | \mathcal{F}_t]$ is a $(\mathcal{D})$-martingale. This decomposition is unique (up to indistinguishability).*

Note that since the conditional expectation is not linear w.r.t. multiplication, we cannot replace, as in the Doob–Meyer decomposition, (2.3) by

$$Z_t = \mathbf{E}\left[\frac{B_\zeta}{B_t}\Big|\mathcal{F}_t\right] \times \mathbf{E}[Z_\zeta | \mathcal{F}_t].$$

Now, we will focus on a new type of decomposition relative to the max operation that plays the role of addition in the Max-Plus semifield $\mathbb{R}_{\max}$: the set of real numbers with the additional point $-\infty$ endowed with the operations $\oplus$ and $\otimes$ such that $x \oplus y = \max(x, y)$ and $x \otimes y = x + y$.

The problem then is to find an optional process $L$ such that

$$Z_t = \mathbf{E}\left[\sup_{t \le u \le \zeta} L_u | \mathcal{F}_t\right] = \mathbf{E}\left[\oint_t^\zeta L_u | \mathcal{F}_t\right].$$

Before that, we first introduce the Max-Plus algebra and describe its basic properties.

2.3. *Max-Plus algebra.* The Max-Plus framework is important for certain problems in discrete mathematics and in computer science applications.

DEFINITION 2.4 (The algebraic structure $\mathbb{R}_{\max}$). The symbol $\mathbb{R}_{\max}$ denotes the set $\mathbb{R} \cup \{-\infty\}$ endowed with the two binary operations $\oplus = \max$ and $\otimes = +$. This algebraic structure $\mathbb{R}_{\max}$ is called the Max-Plus algebra.

It is *an idempotent commutative semifield*, that is, the operation $\oplus$ is associative, commutative and has $0_\oplus = -\infty$ as zero element. The operation $\otimes$ defines a group on $\mathbb{R}$; it is distributive with respect to $\oplus$ and its identity element $e_\otimes$ satisfies $0_\oplus \otimes e_\otimes = -\infty \otimes 0 = 0 \otimes -\infty = -\infty = 0_\oplus$.

If we compare the properties of $\oplus$ and $\otimes$ with those of $+$ and $\times$, we see that:

- we have lost the symmetry of addition [for a given $a$, an element $b$ does not exist such that $a \oplus b = 0_\oplus$, or, equivalently, $\max(b, a) = -\infty$ whenever $a \ne 0_\oplus = -\infty$], but at the same time, we have gained the idempotency of addition: $a \oplus a = a$;
- there are no zero divisors in $\mathbb{R}_{\max}(a \otimes b = 0_\oplus = -\infty \Rightarrow a = 0_\oplus = -\infty$ or $b = 0_\oplus = -\infty)$.

The fact that $\oplus$ is idempotent instead of being invertible is the main original feature of this "exotic" algebra. This property is sufficient for the simplification of formulae and $\mathbb{R}_{max}$ is hence a very effective structure to make algebraic computations [6]. The simplest example of such computations is the resolution of linear equations.

EXAMPLE 2.5 (Linear equation). *Let us examine the solutions in $x$ of*

$$(2.5) \hspace{3cm} z \oplus x = m.$$

*Note that the set of solutions can be empty unlike the classic linear equations. When it is not empty, the set of solutions has a greatest element $x = m$.*

PROOF.   If $m < z$, then there is no solution.

If $m > z$, then $x = m$.

If $m = z$, then any $x \leq z$ is a solution, but $x = m$ is the maximal one.   $\square$

The interested reader is referred to the book by Baccelli, Cohen, Olsder and Quadrat [6] and the references therein. It gives a comprehensive account of deterministic and stochastic Max-Plus linear discrete event systems, together with recent algebraic results (such as symmetrization). The Max-Plus algebra appears to be the right tool to handle synchronization in a *linear* manner, whereas this phenomenon seems to be very nonlinear, or even nonsmooth, "through the glasses" of conventional algebraic tools. Although the initial motivation was essentially found in the study of discrete event systems, it turns out that the theory of *linear* systems on the Max-Plus algebra may be appropriate for other purposes too.

There also exist several excellent survey articles on the subject as the one by Gaubert [28], who presents what he believes to be the minimal core of Max-Plus results, and illustrates these results by typical applications, at the frontier of language theory, control and operations research (performance evaluation of discrete event systems, analysis of Markov decision processes with average cost). The algebraic structure $\mathbb{R}_{max}$ introduces a linear algebra point of view to dynamic programming problems and large deviations. A theory of nonlinear Max-Plus probabilities has been introduced to formalize this point of view (see, e.g., [1] by Akian, [2] by Akian, Quadrat and Viot, [11] by Del Moral and Doisy), where the Max-Plus probability of an event corresponds to the gain of a set of decisions. When the decision space is $\mathbb{R}^+$ and the gain process a deterministic real function, this nonlinear probability is defined as a nonlinear integral:

EXAMPLE 2.6 (Max-Plus integral via a running supremum).   Let $(l_t)_{t \in [0,\zeta]}$ be a real valued function. For any $0 \leq s \leq t \leq \zeta$, the running supremum of $l$ between $s$ and $t$ is by definition a Max-Plus integral on $[s,t]$ w.r.t. $l$,

$$(2.6) \qquad l^*_{s,t} = \sup_{s \le u \le t} l_u = \oint_s^t l_u = I^+([s,t[).$$

The function $t \to l^*_{s,t}$ defined on $[s,\zeta]$ is nondecreasing, right-continuous, provided that $l$ is upper-right semi-continuous (u.r.s.c.), that is, $\limsup_{\varepsilon \downarrow \downarrow 0} l_{t+\varepsilon} \le l_t$.

The Max-Plus additivity property $(\oint_s^t l_v) \oplus (\oint_t^u l_v) = \oint_s^u l_v$ comes from the following relation:

$$\sup_{s \le v \le t} l_v \vee \sup_{t \le v \le u} l_v = \sup_{s \le v \le u} l_v.$$

The process $(l^*_{0,t})$ only increases at points in time $s$ satisfying $l_s = l^*_{0,s}$.

Let $\Omega$ be some "sample space," and $Q$ a function defined on $\Omega$ with negative values and such that $\sup_{\omega \in \Omega} Q(\omega) = 0$. Then for any $A \subset \Omega$, the Max-Plus probability of $A$ is $P^+(A) = \sup_{\omega \in A} Q(\omega)$. $Q(\omega)$ is called the likelihood of $\omega$, and $Q$ the Max-Plus probability density function. A Max-Plus random variable is any $\mathbb{R}_{\max}$-valued function $Z$ on $\Omega$. The Max-Plus expectation of $Z$ is $E^+(Z) = \oint_\Omega Z(\omega) \otimes P^+(d\omega) = \sup_{\omega \in \Omega}[Z(\omega) + Q(\omega)]$. Note that if $E^+(Z) < +\infty$, then $Z$ is Max-Plus integrable. It is also immediate that $E^+$ is stable by increasing limit and linear under Max-Plus addition and scalar multiplication.

In the particular case where $\Omega = \mathbb{R}^+$ and $Z(\omega) = l_{]s,t]}(\omega)$ where $l_{]s,t]}$ stands for the Max-Plus indicator function which equals $0$ if $\omega \in ]s,t]$ and $-\infty$ otherwise, $E^+(Z)$ comes down to our Max-Plus integral (2.6).

The theory of the Max-Plus probability calculus has been deeply developed by W. H. Fleming who has considered the notion of Max-Plus additive integral in his paper [25]. This paper is also concerned with a concept of Max-Plus martingale, which is similar to the concept of exponential-maxingale as defined in [41]. The results presented there were intended as initial steps toward a Max-Plus stochastic calculus (review of some aspects of Max-Plus probability, concept of Max-Plus stochastic differential equation and associated backward and forward partial differential equations, etc.).

2.4. *Main result.* We here give the main theorem of the paper, in a similar form to that of Theorem 2.2. The first part is the Max-Plus version of the supermartingale-decomposition, expressed in terms of the running supremum of some process $L$. In contrast with what happens in the additive or multiplicative decomposition [equations (2.2) and (2.4)], here the identity $\mathbf{E}[\Lambda_T | \mathcal{F}_t] = Z_t \oplus \Lambda_t$ does not hold true at any time $t$ in $[0,\zeta]$. We just have an inequality.

We now formulate the main theorem of the paper. The proof of existence of a process satisfying the properties below is relegated to Section 4. It is established in the case where $Z$ is quasi-left-continuous. Let us first focus on the question of uniqueness.

THEOREM 2.7.   *Let $Z$ be a $(\mathcal{D})$-supermartingale defined on $[0, \zeta]$, quasi-left-continuous and assume that the filtration $\{\mathcal{F}_t\}$ is quasi-left-continuous.*

1. *$Z$ admits the following Max-Plus decomposition:*

$$(2.7) \quad Z_t = \mathbf{E}\left[\sup_{t \le u \le \zeta} L_u \vee Z_\zeta | \mathcal{F}_t\right] = \mathbf{E}\left[\oint_{[t,\zeta]} L_u \oplus Z_\zeta | \mathcal{F}_t\right], \qquad 0 \le t \le \zeta,$$

   *where $L = (L_t)_{0 \le t \le \zeta}$ is an optional upper-right semi-continuous process, satisfying $L_\zeta \le Z_\zeta$.*

2. *Let $L_{t,s}^*$ be the càdlàg running supremum of $L_u$: $L_{t,s}^* = \sup_{t \le u \le s} L_u = \oint_{[t,s]} L_u$.*

   *Define the $(\mathcal{D})$-martingale $M^\oplus$ by*

$$(2.8) \qquad\qquad M_t^\oplus := \mathbf{E}[L_{0,\zeta}^* \vee Z_\zeta | \mathcal{F}_t] \qquad \forall t \in [0, \zeta].$$

$$(2.9) \qquad Then \qquad M^\oplus \ge \max(Z, L_{0,\cdot}^*) = Z \oplus L_{0,\cdot}^*,$$

   *and at any stopping time $S \le \zeta$,*

$$(2.10) \qquad L_S = L_{0,S}^* \Rightarrow M_S^\oplus = Z_S \oplus L_{0,S}^* = Z_S.$$

   *In particular, the following "flat-off" condition holds true:*

$$(2.11) \qquad\qquad \int_{[0,\zeta]} (M_s^\oplus - Z_s) \, dL_{0,s}^* = 0.$$

3. *This decomposition is unique in the sense of Theorem 2.8 below.*

PARTIAL PROOF.   1. See Section 4 for the construction of the process $L_u$.
2. For any time $t$ in $[0, \zeta]$, by the $\oplus$ additivity of $L^*$, we have that

$$M_t^\oplus = \mathbf{E}[L_{0,\zeta}^* \vee Z_\zeta | \mathcal{F}_t] = \mathbf{E}[L_{0,t}^* \vee L_{t,\zeta}^* \vee Z_\zeta | \mathcal{F}_t]$$
$$\ge L_{0,t}^* \vee \mathbf{E}[L_{t,\zeta}^* \vee Z_\zeta | \mathcal{F}_t] = L_{0,t}^* \vee Z_t.$$

Moreover, if $S$ is a stopping time at which $L_S = L_{0,S}^*$, then $L_{S,\zeta}^* = L_{0,\zeta}^*$ and so $Z_S = M_S^\oplus$.   $\square$

The above representation (2.7) provides an additive decomposition for supermartingales in the algebraic structure $\mathbb{R}_{\max}$, analogous to the Doob–Meyer's one, but here the nondecreasing process is only optional and not necessarily predictable. This restriction is similar to that appearing in the Kramkov decomposition [34].

We have also to note that, contrary to what occurs with the Max-Plus decomposition, in the preceding decompositions the equality $M_S^A = Z_S$, where $M^A$ is the martingale of the additive Doob–Meyer decomposition (resp. $M_S^P = Z_S$, where $M^P$ is the martingale of the multiplicative decomposition) only holds before the first stopping time at which the nondecreasing process $A$ (resp. $B$) increases.

2.4.1. *Uniqueness result.* For the uniqueness, we have not to assume a priori that the nondecreasing process, involved in the Max-Plus decomposition, is a running supremum. We only use the "flat-off" condition (2.11).

THEOREM 2.8 (Uniqueness). *Let $\{\mathcal{F}_t\}$ be a quasi-left-continuous filtration and $Z$ a $(\mathcal{D})$-supermartingale. Assume there exists a $(\mathcal{D})$-martingale $M$ with $M_0 = Z_0$, and a càdlàg adapted nondecreasing process $\Lambda$ taking values in $[-\infty, +\infty)$, such that, almost surely,*

$$(2.12) \qquad M_t \geq Z_t \qquad \forall t \in [0, \zeta], \qquad M_\zeta = \Lambda_\zeta \vee Z_\zeta.$$

*We further assume that $\Lambda$ only increases at times $t \leq \zeta$ when $M_t = Z_t$, that is $\Lambda$, satisfies the "flat-off condition"*

$$(2.13) \qquad \int_{[0,\zeta]} (M_t - Z_t) \, d\Lambda_t = 0 \qquad a.s.$$

*Then the martingale $M$ is unique and denoted by $M^\oplus$ in the sequel.*

*In addition, given such a martingale $M^\oplus$, the set $\mathcal{K}$ of the nondecreasing processes $\Lambda$ satisfying (2.12) and (2.13) has a maximal element, which also satisfies condition (2.13). It is denoted by $\Lambda^{\max}$.*

REMARK 2.9. (a) When $\Lambda$ is a running supremum $L_{0,t}^*$, $\Lambda$ only increases when $L_{0,t}^* = L_t$. So, the "flat-off condition" implies that condition (2.10) holds true.

(b) At maturity, "the flat-off condition" means that on the subset $\{\Lambda_\zeta > \Lambda_{\zeta-\epsilon}, \forall \epsilon > 0\}$, $\Lambda_\zeta \leq Z_\zeta$. This inequality also holds true if $\Lambda_\zeta^+ > \Lambda_\zeta$, where $\Lambda_\zeta^+ = \Lambda_\zeta \vee Z_\zeta$.

(c) Observe that if $Z$ is bounded by below by a real $c$, $M^\oplus$ is also bounded by below by $c$, and the nondecreasing process $\Lambda \oplus c$ satisfies conditions (2.9) and (2.10).

REMARK 2.10 (Dynamic programming). Unlike the other supermartingale decompositions where, for any stopping time $\tau$, the decomposition on the time interval $[0, \zeta]$ is also the decomposition on $[0, \tau]$, here this property only holds true if $M_\tau^\oplus = Z_\tau \oplus \Lambda_\tau$.

Before establishing the uniqueness result, let us give some examples associated with degenerated supermartingales, first as a decreasing process and then as a martingale.

EXAMPLE 2.11. (a) *Case of a nonincreasing process*: if $Z$ is an adapted integrable nonincreasing process, then, for each $t \in [0, \zeta]$,

$$M_t^{\oplus} = \Lambda_t^{\max} = Z_0 \qquad \text{a.s.}$$

In fact, it can be easily seen that $Z_0$ is a martingale that satisfies the Max-Plus decomposition. It is also the only one since the martingale involved in the decomposition is unique.

(b) *Case of a martingale*: if $Z$ is a martingale, then, for each $t \in [0, \zeta]$,

$$M_t^{\oplus} = Z_t \quad \text{and} \quad \Lambda_t^{\max} \le Z_t = \mathbf{E}[Z_\zeta | \mathcal{F}_t].$$

Hence, $\Lambda^{\max}$ is the greatest càdlàg nondecreasing process dominated by $Z$, and defines the conditional infimum of the random variable $Z_\zeta$ with respect to the $\sigma$-algebra $\mathcal{F}_t$:

$$\Lambda_t^{\max} = \mathcal{F}_t - \operatorname{ess\,inf}\{Z_\zeta\} = \operatorname{ess\,sup}\{Y_t \in \mathcal{F}_t | Y_t \le Z_\zeta \text{ a.s.}\} \qquad \forall t \in [0, \zeta].$$

In fact, for any $r \in \mathbb{Q}$, where $\mathbb{Q}$ denotes the field of rationals, let us define $\Lambda_r^{\max}$ as the conditional infimum of $Z_\zeta$ with respect to $\mathcal{F}_r$. The sequence $(\Lambda_r^{\max})_{r \in \mathbb{Q}}$ is nondecreasing with right-continuous regularization $\Lambda_t^{\max}$. It is clear that $\Lambda_t^{\max} \le \mathbf{E}(Z_\zeta | \mathcal{F}_t) = Z_t$. Moreover, if $\Lambda$ is a càdlàg nondecreasing process such that $\Lambda_t \le \Lambda_\zeta \le Z_\zeta$, then $\Lambda_t \le \Lambda_t^{\max}$.

REMARK 2.12. The conditional infimum is defined in Barron, Cardialaguet and Jensen [9]. The concept of maxingale is also introduced, as associated with the conditional infimum with respect to a filtration, and used to develop the new theory of optimal stopping in $L^\infty$, as well as the concept of an *absolutely* optimal stopping time. In our case, $\Lambda^{\max}$ defines a maxingale since $\Lambda_t^{\max} = \mathcal{F}_t - \operatorname{ess\,inf}\{\Lambda_\zeta^{\max}\}$, a.s. for each $t \in [0, \zeta]$. In contrast to martingales, it turns out that maxingales are easier to analyze since they basically always converge. It should be also noted that these processes are different from those appearing in the theory of Fleming or Puhalskii [25, 41], since it is not the probability Max-Plus which is used here but the usual one.

The proof of uniqueness is essentially based on the observation that $M_\zeta$ is equal to the terminal value $\Lambda_\zeta^+$ of a nondecreasing process. This kind of argument was first introduced by El Karoui and Jeanblanc-Picqué in [16]. It also appears in the papers [17] by El Karoui, Jeanblanc-Picqué and Lacoste and [7] by Bank and El Karoui.

PROOF OF THEOREM 2.8.   In the proof, we use the short notation $\Lambda^+$ introduced in Remark 2.9 for the process $\Lambda$ jumping at time $\zeta$, such that $\Lambda_\zeta^+ = \Lambda_\zeta \vee Z_\zeta$. $\Lambda^+$ still satisfies the flat-off condition.

Assume that there exists two decompositions $(M^1, \Lambda^{+,1})$ and $(M^2, \Lambda^{+,2})$ satisfying the previous conditions (2.12) and (2.13). For the sake of simplicity, we first suppose that $\Lambda_0^{+,1}$ and $\Lambda_0^{+,2}$ are finite (note that this assumption is automatically satisfied if $Z$ is bounded by below). Then at the end of the proof, we will show how this artificial assumption can be relaxed.

(a) Let $f$ be an arbitrary positive, regular, convex function in $\mathcal{C}_b^2$, null in zero [e.g., the $\mathcal{C}^2$-bounded regularization of $x \mapsto x^+$ on the intervals of the form $[-\epsilon, \epsilon]$ and $(-\infty, -\frac{1}{\epsilon}] \cup [\frac{1}{\epsilon}, \infty)(\varepsilon > 0)]$. Since $f(0) = 0$, the convexity of $f$ yields

$$f(M_\zeta^1 - M_\zeta^2) \leq f'(M_\zeta^1 - M_\zeta^2)(M_\zeta^1 - M_\zeta^2) = f'(\Lambda_\zeta^{+,1} - \Lambda_\zeta^{+,2})(M_\zeta^1 - M_\zeta^2).$$

We use the classic differential rule for finite variation processes; it is convenient to introduce the discrete derivative of $f'$, $f_d''$ as

$$\begin{cases} f_d''(x, \delta) = \dfrac{1}{\delta}(f'(x + \delta) - f'(x)), & \text{if } \delta \neq 0, \\ f_d''(x, 0) = f''(x), & \text{if } \delta = 0. \end{cases}$$

In the sequel, we set $\Delta_s^{1,2} = (\Lambda_s^{+,1} - \Lambda_s^{+,2}) - (\Lambda_{s^-}^{+,1} - \Lambda_{s^-}^{+,2})$ for $s$ in $(0, \zeta]$. Thus,

$$\begin{aligned} f'(\Lambda_\zeta^{+,1} - \Lambda_\zeta^{+,2}) = {} & f'(\Lambda_0^{+,1} - \Lambda_0^{+,2}) \\ & + \int_{(0,\zeta]} f_d''(\Lambda_{s^-}^{+,1} - \Lambda_{s^-}^{+,2}, \Delta_s^{1,2}) \, d(\Lambda_s^{+,1} - \Lambda_s^{+,2}). \end{aligned}$$

Since $f$ is a convex regular function in $\mathcal{C}_b^2$, $f_d''$ is positive and bounded. As we do not really need the explicit form of $f_d''$, we introduce the short notation $\tilde{f}_d''(s) := f_d''(\Lambda_{s^-}^{+,1} - \Lambda_{s^-}^{+,2}, \Delta_s^{1,2})$.

Moreover, note that the current value of the uniformly integrable martingale $M^1 - M^2$ at any time $s$ is nothing else but the conditional expectation of its terminal value w.r.t. the filtration $\mathcal{F}_s$, whence

$$\begin{aligned} & \mathbf{E}[f'(\Lambda_\zeta^{+,1} - \Lambda_\zeta^{+,2})(M_\zeta^1 - M_\zeta^2)] \\ & \quad = \mathbf{E}[f'(\Lambda_0^{+,1} - \Lambda_0^{+,2})(M_\zeta^1 - M_\zeta^2)] \\ & \qquad + \mathbf{E}\left[\int_{(0,\zeta]} (M_s^1 - M_s^2)\tilde{f}_d''(s) \, d(\Lambda_s^{+,1} - \Lambda_s^{+,2})\right]. \end{aligned}$$

In addition, $\Lambda^{+,1}$ (resp. $\Lambda^{+,2}$) only increases at times $t \leq \zeta$ when $M_t^1 = Z_t$ (resp. $M_t^2 = Z_t$), whence

$$\int_{(0,\zeta]} (M_s^1 - M_s^2)\tilde{f}_d''(s) \, d(\Lambda_s^{+,1} - \Lambda_s^{+,2})$$

$$= \int_{(0,\zeta]} \tilde{f}_d''(s)((Z_s - M_s^2) \, d\Lambda_s^{+,1} - (M_s^1 - Z_s) \, d\Lambda_s^{+,2}) \leq 0.$$

Thus, these considerations lead to

$$\mathbf{E}[f'(\Lambda_\zeta^{+,1} - \Lambda_\zeta^{+,2})(M_\zeta^1 - M_\zeta^2)] \leq \mathbf{E}[f'(\Lambda_0^{+,1} - \Lambda_0^{+,2})(M_0^1 - M_0^2)] = 0,$$

whence $\mathbf{E}[f(M_\zeta^1 - M_\zeta^2)] \leq 0$ for all convex functions in $\mathcal{C}_b^2$, and the desired result follows at once.

(b) In the case where $\Lambda_0^1$ and $\Lambda_0^2$ are infinite, we just have to set $-\infty + \infty = 0$.

(c) Let us consider the set of nondecreasing processes in $\mathcal{K}$ satisfying

$$\int_{[0,\zeta]} (M_t^\oplus - Z_t) \, d\Lambda_t^+ = 0, \qquad M_\zeta^\oplus = \Lambda_\zeta^+ = \Lambda_\zeta \vee Z_\zeta.$$

This set is stable by Max-Plus addition. In fact, if $M_\zeta^\oplus = \Lambda_\zeta^1 \vee Z_\zeta = \Lambda_\zeta^2 \vee Z_\zeta$, the same equality holds for $\Lambda^1 \vee \Lambda^2$ as it has been shown in Example 2.5 on the linear equation (2.5) in the Max-Plus algebra. Moreover, since $\Lambda^{+,1}$ and $\Lambda^{+,2}$ only increase at points in time $t$ when $M_t^\oplus = Z_t$, $\Lambda^{+,1} \vee \Lambda^{+,2}$ satisfies the same property.

The idea now is to introduce the essential supremum of this family of nondecreasing processes. To do so, we consider for any $r \in \mathbb{Q}^+$, $\widetilde{\Lambda_r} = \operatorname{ess\,sup}\{\Lambda_r; \Lambda \in \mathcal{K}\}$. The family $(\widetilde{\Lambda_r})_{r \in \mathbb{Q}^+}$ is clearly nondecreasing, and for any real $t$ in $[0,\zeta)$, we define by $\Lambda_t^{\max} = \lim_{r \downarrow\downarrow t} \widetilde{\Lambda_r}$, the right-regularization of $\widetilde{\Lambda_r}$. □

## 3. Generalization of Darling–Liggett–Taylor's point of view on American options. 
Mathematical finance has popularized a particular kind of optimal stopping problems, called American options.

In this section we generalize the ideas of Darling, Liggett and Taylor, who considered in their paper [10] American Call options written on partial sums $S_n$ of independent and identically distributed random variables with negative drift, and characterized optimal stopping times in terms of the running supremum of the underlying $S_n$.

3.1. *Max-Plus decomposition and American Call options.* Hence, based on a supermartingale representation in terms of a running supremum of some index process $L_t$, we characterize an optimal stopping time in terms of $L$ and represent the value function using the running supremum of Bank and Föllmer that have provided in [8] the same kind of representation based on another decomposition (see Remark 3.7).

THEOREM 3.1. *Let $Z$ be a supermartingale of class $(\mathcal{D})$ with the following Max-Plus decomposition in terms of an upper-right semi-continuous* process $L_t$: $Z_t = \mathbf{E}[\sup_{t \leq s \leq \zeta} L_s \vee Z_\zeta | \mathcal{F}_t] = \mathbf{E}[L_{t,\zeta}^* \oplus Z_\zeta | \mathcal{F}_t]$, where the filtration $(\mathcal{F}_t)$ is assumed to be quasi-left-continuous.

The problem to find an optimal stopping time $\tau_t^*(m)$ to the American Call option $C^{\mathrm{Am}}(Z, m)$ written on $Z$, and with strike $m$,

$$C_t^{\mathrm{Am}}(Z, m) = \operatorname{ess\,sup}_{t \leq s \leq \zeta} \mathbf{E}[(Z_s - m)^+ | \mathcal{F}_t] = \mathbf{E}[(Z_{\tau_t^*(m)} - m)^+ | \mathcal{F}_t],$$

has an explicit universal solution (w.r.t. the strike $m \geq 0$) given by

(3.1)
$$\tau_t^*(m) = T_t(m) \wedge \zeta := \inf\{s \geq t; L_{t,s}^* \geq m\} \wedge \zeta$$
$$= \inf\{s \geq t; L_s \geq m\} \wedge \zeta.$$

Moreover, the American Call option $C^{\mathrm{Am}}(Z, m)$ is explicitly characterized as a lookback one:

(3.2)
$$C_t^{\mathrm{Am}}(Z, m) = \mathbf{E}[(L_{t,\zeta}^* \vee Z_\zeta - m)^+ | \mathcal{F}_t]$$
$$= \mathbf{E}\left[\left(\sup_{t \leq s \leq \zeta} L_s \vee Z_\zeta - m\right)^+ \Big| \mathcal{F}_t\right], \qquad t \leq \zeta.$$

Hence thanks to equation (3.1) which links the optimal stopping time $T_t(m)$ to the index process $L_t$, we do not need to compute the value function in order to solve the optimal stopping problem, if we have the Max-Plus decomposition of $Z$.

The key point is that we can omit the conditioning by $\mathcal{F}_S$ in the Max-Plus representation of $Z_S$ and thus replace $Z$ by $L_{\cdot,\zeta}^*$, when computing the American Call price.

PROOF OF THEOREM 3.1. First note that the process $(\mathbf{E}[(L_{t,\zeta}^* \vee Z_\zeta - m)^+ | \mathcal{F}_t])_{t \in [0,\zeta]}$ defines a supermartingale since it is the conditional expectation of a nonincreasing process. Moreover, since $Z_S = \mathbf{E}[L_{S,\zeta}^* \vee Z_\zeta | \mathcal{F}_S]$ for any stopping time $S \leq \zeta$, Jensen's inequality implies

$$\mathbf{E}[(L_{S,\zeta}^* \vee Z_\zeta - m)^+ | \mathcal{F}_S] \geq (\mathbf{E}[L_{S,\zeta}^* \vee Z_\zeta | \mathcal{F}_S] - m)^+ = (Z_S - m)^+,$$

for any stopping time $S \leq \zeta$. Thus, the supermartingale $(\mathbf{E}[(L_{t,\zeta}^* \vee Z_\zeta - m)^+ | \mathcal{F}_t])_{t \in [0,\zeta]}$ dominates the process $(Z_t - m)^+$, and necessarily its Snell envelope $C_t^{\mathrm{Am}}(Z, m)$ as well.

Define $T_t(m)$ as the first stopping time after $t$, at which the process $(L_{t,u}^*)_{u \geq t}$ goes beyond $m$:

$$T_t(m) := \inf\{s \in [t, \zeta]; L_{t,s}^* \geq m\}, \qquad = \infty^+ \text{ if the set is empty.}$$

On the set $\{T_t(m) \leq \zeta\}$, $L^*_{T_t(m),\zeta} \vee Z_\zeta \geq m$ and, hence, $Z_{T_t(m)} = E[L^*_{T_t(m),\zeta} \vee Z_\zeta | \mathcal{F}_{T_t(m)}] \geq m$. This yields

$$\mathbf{E}[(L^*_{t,\zeta} \vee Z_\zeta - m)^+ | \mathcal{F}_t] = \mathbf{E}[\mathbf{E}[\mathbf{1}_{\{T_t(m) \leq \zeta\}}(L^*_{T_t(m),\zeta} \vee Z_\zeta - m)^+ | \mathcal{F}_{T_t(m)}] | \mathcal{F}_t]$$
$$= \mathbf{E}[\mathbf{1}_{\{T_t(m) \leq \zeta\}} \mathbf{E}[L^*_{T_t(m),\zeta} \vee Z_\zeta - m | \mathcal{F}_{T_t(m)}] | \mathcal{F}_t]$$
$$= \mathbf{E}[\mathbf{1}_{\{T_t(m) \leq \zeta\}}(Z_{T_t(m)} - m)^+ | \mathcal{F}_t]$$
$$= \mathbf{E}[(Z_{T_t(m) \wedge \zeta} - m)^+ | \mathcal{F}_t], \qquad \text{a.s.,}$$

where the last equality follows from the fact that $Z_\zeta(m) > Z_\zeta$ on the set $\{T_t(m) = \infty^+\}$, and since $Z_\zeta(m) = Z_\zeta \vee m$, it comes that $Z_\zeta < m$.

Hence, $\mathbf{E}[(L^*_{t,\zeta} \vee Z_\zeta - m)^+ | \mathcal{F}_t] = \mathbf{E}[(Z_{T_t(m) \wedge \zeta} - m)^+ | \mathcal{F}_t] \leq C^{\mathrm{Am}}_t(Z, m)$ by definition of the Snell envelope. This completes the proof since we have already shown the reverse inequality.  $\square$

Supermartingale processes $Z$ with independent and stationary additive or multiplicative increments give us nice examples where the Max-Plus decomposition is obvious. We hereafter give some illustrative examples which show that we can completely solve the American Call problem with no need to the explicit price. We just have to compute the expectation of the running supremum of $Z$, and not its probability distribution. Similar ideas are used by Darling, Liggett and Taylor [10] to solve the optimal stopping problem in the discrete case.

### 3.2. *Positive multiplicative Lévy processes.*

Running suprema of Lévy processes have been, in particular, studied in [4] by Asmussen, Avram and Pistorius, who have derived explicit formulae for the pricing of Russian and perpetual American put options, under exponential phase-type Lévy models.

Let $Z$ be a positive multiplicative Lévy process with initial value $x$ and such that $\mathbf{E}[\sup_{0 \leq t \leq \zeta} Z_t] < +\infty$. Since $Z_0 = x$, it comes that $Z_t = x\mathcal{Z}_t$, where $\mathcal{Z}$ is the process that equals $Z$ when $Z_0 = 1$. We first assume the maturity $\zeta$ to be infinite.

### 3.2.1. *Infinite horizon.*

Thanks to the independence property of the relative increments of $Z$, we can easily determine a nondecreasing process $L^*_{0,t} = \sup_{0 \leq u \leq t} L_u$, satisfying the Max-Plus decomposition of $Z$. Then applying Theorem 3.1, we immediately obtain an explicit characterization of an optimal stopping time without computing the American Call price. The only quantity we need to calculate is $x\mathbf{E}[\sup_{t \geq 0} Z_t]^{-1} = x\mathbf{E}[Z^*_{0,\infty}]^{-1} = \mathbf{E}[\mathcal{Z}^*_{0,\infty}]^{-1}$, which does not depend on the initial value $x$. We assume in the following proposition that $(\mathcal{F}_t)$ is a quasi-left-continuous filtration.

PROPOSITION 3.2. *Let $Z$ be a positive multiplicative Lévy process defining a supermartingale with initial value $x$ and such that $\mathbf{E}[\sup_{0 \leq t \leq \infty} Z_t] < +\infty$. Then using the notation $Z_t^* := Z_{0,t}^* = \sup_{0 \leq u \leq t} Z_u$, we get:*

1. $Z_t = b\mathbf{E}[Z_{t,\infty}^*|\mathcal{F}_t]$  *and*  $L_{0,t}^* = bZ_t^*$  *where* $b = \dfrac{1}{\mathbf{E}[\mathcal{Z}_{0,\infty}^*]}$.

2. *An exercise boundary to the perpetual American Call option $C^{\mathrm{Am}}(Z, m)$ is given by $E^{\mathrm{c}}(m) = m\mathbf{E}[\mathcal{Z}_{0,\infty}^*]$ and the corresponding optimal stopping time is explicitly characterized by*

$$T_t(m) = \inf\{s \geq t; bZ_s \geq m\} = \inf\{s \geq t; Z_s \geq m\mathbf{E}[\mathcal{Z}_{0,\infty}^*]\}.$$

PROOF.  1. We use the previous notation $Z_{s,t}^* = \sup_{s \leq u \leq t} Z_u$. Thanks to the independence of relative increments and the integrability property of $Z_{0,\infty}^*$, we have that

$$(3.3) \quad Z_t = b\mathbf{E}[Z_{t,\infty}^*|\mathcal{F}_t], \qquad \frac{1}{b} = \frac{1}{x}\mathbf{E}\Big[\sup_{0 \leq u} Z_u\Big] = \mathbf{E}[\mathcal{Z}_{0,\infty}^*], \qquad L_t = bZ_t,$$

since

$$\mathbf{E}[Z_{t,\infty}^*|\mathcal{F}_t] = Z_t\mathbf{E}\Big[\sup_{t \leq u} \frac{Z_u}{Z_t}\Big|\mathcal{F}_t\Big]$$

$$= Z_t\mathbf{E}\Big[\sup_{0 \leq u}\Big(\frac{Z_{u+t}}{Z_t}\Big)\Big|\mathcal{F}_t\Big] = Z_t\mathbf{E}\Big[\sup_{0 \leq u} \mathcal{Z}_u\Big]. \qquad \square$$

We also note that, in light of Theorem 3.1, a perpetual American Call written on $Z$ with strike $m$ is nothing else but a perpetual lookback option written on the index process $L$ and with the same strike $m$:

$$(3.4) \quad C_0^{\mathrm{Am}}(Z, m) = \mathbf{E}[(bZ_{0,\infty}^* - m)^+] = \mathbf{E}\Big[\Big(\frac{Z_{0,\infty}^*}{\mathbf{E}[\mathcal{Z}_{0,\infty}^*]} - m\Big)^+\Big].$$

REMARK 3.3.  When the time horizon is finite and denoted by $T$, the constant $b$ in the Max-Plus decomposition of $Z$ is replaced by a function $b(\cdot)$ such that, at any time $t$,

$$Z_t = \mathbf{E}\Big[\sup_{t \leq u \leq T} b(T - u)Z_u|\mathcal{F}_t\Big].$$

However, we do not have an explicit formula for the function $b(\cdot)$.

Let us give some examples of positive multiplicative Lévy processes $Z$ for which the corresponding perpetual American Call option can be solved explicitly. Since we only need to know the exercise boundary, the following examples come down to calculate the expectation of the maximum of $Z$. We start with the geometric Brownian motion motivated by financial applications.

*Examples of calculations of the expectation of the maximum Geometric Brownian motion.*  Let us consider a geometric Brownian motion with negative drift ($r \geq 0$) and positive initial value, so as to be a supermartingale:

$$\frac{dZ_t}{Z_t} = -r\,dt + \sigma\,dW_t, \qquad Z_0 = x > 0.$$

Set $\gamma = 1 + \frac{2r}{\sigma^2}$. It is well known that $\mathbf{P}[Z_{0,\infty}^* \geq m] = (\frac{x}{m} \wedge 1)^\gamma$, $\mathbf{E}[Z_{0,\infty}^*] = \frac{\gamma}{\gamma-1}x$ and, thus, $\mathbf{E}[\mathcal{Z}_{0,\infty}^*] = \frac{\gamma}{\gamma-1}$. This classic result is proven in Section 5.

*Geometric Lévy process.*  The same decomposition holds for a jumping *geometric Lévy process* $Z$ that defines a supermartingale. However, we have to introduce other assumptions so as to satisfy the condition $\mathbf{E}[Z_{0,\infty}^*] < \infty$. We follow hereafter the notation of Mordecki in his paper [38].

Let us consider the particular case where the horizon $\zeta$ is infinite and $Z_t = xe^{X_t}$, where $X$ defines an upper semi-continuous process, or equivalently, a Lévy process with no positive jumps:

$$\mathbf{E}e^{i\mu X_t} = \exp\left\{t\left[i\mu a - \tfrac{1}{2}\sigma^2\mu^2 + \int_{-\infty}^0 (e^{i\mu y} - 1 - i\mu y\mathbf{1}_{\{-1 < y < 0\}})\Pi(dy)\right]\right\},$$

with $a$ and $\sigma \geq 0$ two real constants, and $\Pi$ a positive measure supported on the set $(-\infty, 0)$ such that $\int(1 \wedge y^2)\Pi(dy) < +\infty$. The Laplace exponent is defined for $\lambda \geq 0$ by

$$\kappa(\lambda) = a\lambda + \tfrac{1}{2}\sigma^2\lambda^2 + \int_{-\infty}^0 (e^{\lambda y} - 1 - \lambda y\mathbf{1}_{\{-1 < y < 0\}})\Pi(dy),$$

and satisfies $\mathbf{E}(e^{\lambda X_t}) = e^{t\kappa(\lambda)}$. We also assume that the process $e^{rt}e^{X_t}$ defines a martingale, which implies that $\mathbf{E}[e^{X_t}] = e^{-rt}$, that is, $\kappa(1) = -r$.

Since $\kappa(0) = 0$, $\kappa$ is convex and $\lim_{\lambda \to +\infty} \kappa(\lambda) = +\infty$, there exists $\gamma_{\text{Lévy}} > 1$ such that $\kappa(\gamma_{\text{Lévy}}) = 0$. Hence, in light of the paper [38], we deduce that

$$\mathbf{E}[\mathcal{Z}_{0,\infty}^*] = \frac{\gamma_{\text{Lévy}}}{\gamma_{\text{Lévy}} - 1} \qquad \text{where } \gamma_{\text{Lévy}} > 1 \text{ is such that } \kappa(\gamma_{\text{Lévy}}) = 0.$$

*Closed formulae.*  Numerically, thanks to equation (3.4), we can compute the price of the perpetual American Call $C_0^{\text{Am}}(Z, m)$ with simply two Monte-Carlo simulations, one for $\mathbf{E}[\mathcal{Z}_{0,\infty}^*]$ and the other for the whole expectation. We just have to check beforehand that $\mathbf{E}[Z_{0,\infty}^*] < +\infty$.

However, in many cases of Lévy processes, these expectations are easy to calculate and can be got in an explicit form. For example, the next proposition gives the explicit expression of $C_t^{\text{Am}}(Z, m)$ in the case where $Z$ evolves according to a geometric Brownian motion, but since this price is needless to solve the optimal stopping problem, we relegate the proof to Section 5.

PROPOSITION 3.4. *Let us assume the filtration $(\mathcal{F}_t)$ to be quasi-left-continuous and consider a geometric Brownian motion $Z$ with parameters $(-r, \sigma)$. The price of the American Call option written on the underlying $Z$, with strike $m$ is given by*

$$C_t^{\mathrm{Am}}(Z, m) = \mathbf{E}\left[\left(\frac{\gamma - 1}{\gamma} Z_{t,\infty}^* - m\right)^+ \Big| \mathcal{F}_t\right]$$

$$= \begin{cases} \left(\dfrac{m}{\gamma - 1}\right)^{1-\gamma} \left(\dfrac{Z_t}{\gamma}\right)^\gamma, & \text{if } \dfrac{\gamma - 1}{\gamma} Z_t \leq m, \\ Z_t - m, & \text{otherwise.} \end{cases}$$

*American Put options.* Note that the American Call option $C^{\mathrm{Am}}(Z, m)$ with no discount factor comes down to a "classic" American Put by a simple change of probability measure.

In fact, using the martingale property of the positive continuous process $e^{rt} Z_t$, we define a new equivalent probability measure $\mathbf{Q}^Z$ on $\mathcal{F}_t$ by its Radon–Nikodym density with respect to $\mathbf{P}$:

$$\frac{d\mathbf{Q}^Z}{d\mathbf{P}}\bigg|_{\mathcal{F}_t} = e^{rt}\frac{Z_t}{Z_0} = e^{rt}\frac{Z_t}{x}.$$

Then taking $xZ^{-1}$ as a new numéraire, we transform our American Call option into a classic American Put option with new underlying $mxZ^{-1}$ and new strike $x$:

$$\begin{aligned}(3.5) \qquad C_0^{\mathrm{Am}}(Z, m) &= \sup_\tau \mathbf{E}_{\mathbf{P}}[(Z_\tau - m)^+] \\ &= \sup_\tau \mathbf{E}_{\mathbf{Q}^Z}[xe^{-r\tau}Z_\tau^{-1}(Z_\tau - m)^+] \\ &= \sup_\tau \mathbf{E}_{\mathbf{Q}^Z}[e^{-r\tau}(x - mxZ_\tau^{-1})^+] \\ &= mx\mathrm{Put}_0^{\mathrm{Am}}(x^{-1}, m^{-1}).\end{aligned}$$

Observe that the new underlying $Z^{-1}$ defines a submartingale under the probability measure $\mathbf{Q}^Z$ and evolves according to a positive multiplicative Lévy process. Moreover, the exercise boundary $E^p(m^{-1})$ of this new "classic" Put option is nothing else but the inverse of $E^c(m)$, that is, $E^p(m^{-1}) = \frac{\gamma - 1}{\gamma} m^{-1}$.

This Put-Call duality formula (3.5), based on an adequate change of probability measure, can be more generally extended to one-dimensional time homogeneous diffusion processes with a volatility function. However, the two "dual" underlyings will not necessarily have the same dynamics any more. We refer the reader to [24] for more details about the duality between prices of Put and Call options.

In the case where $Z = xe^{X_t}$, where $X$ defines a Lévy process with no positive jumps, it can be shown that, by an adequate change of probability, $C_t^{\mathrm{Am}}(Z, m)$ is nothing else but the price of a perpetual Put option, written on the underlying asset $me^{-X_t}$ and with a strike $K = x$. Based on [38], we deduce a closed formula for $C_t^{\mathrm{Am}}(Z, m)$ as a function of $Z_t$. This function is the same as in Proposition 3.4, but with the parameter $\gamma_{\text{Lévy}}$.

3.2.2. *Independent exponential r.v. horizon.* Assume now that the time horizon $\zeta$ defines an independent exponential r.v. with parameter $\beta > 0$. Let $\mathcal{G}_t$ be the right-continuous augmented filtration generated by $\mathcal{F}_{t \wedge \zeta} \vee \sigma(t \wedge \zeta)$. Then one can observe that, on the set $\{t < \zeta\}$, any $\mathcal{G}_t$-measurable r.v. on $\Omega \times \mathbb{R}^+$ is also $\mathcal{F}_t$-measurable. So, for any r.v. $X \in \mathcal{G}_\zeta$,

$$(3.6) \quad \mathbf{E}[X|\mathcal{G}_t]\mathbf{1}_{\{t<\zeta\}} = \frac{\mathbf{E}[X\mathbf{1}_{\{t<\zeta\}}|\mathcal{F}_t]}{\mathbf{E}[\mathbf{1}_{\{t<\zeta\}}|\mathcal{F}_t]}\mathbf{1}_{\{t<\zeta\}} = e^{\beta t}\mathbf{E}[X\mathbf{1}_{\{t<\zeta\}}|\mathcal{F}_t]\mathbf{1}_{\{t<\zeta\}}.$$

In order to derive an explicit decomposition of $Z$ in the Max-Plus algebra relative to the $\mathcal{G}$-filtration, we first observe that, on the set $\{t < \zeta\}$, $\frac{Z_{t,\zeta}^*}{Z_t}$ is conditionally independent of $Z_t$ given $\mathcal{G}_t$ and has the same distribution as $\frac{Z_{0,\zeta}^*}{x}$. We hence obtain

$$(3.7) \quad \begin{aligned} Z_t\mathbf{1}_{\{t<\zeta\}} &= \mathbf{E}\left[\int_t^\infty \beta e^{-\beta(s-t)}\frac{Z_{t,s}^*}{Z_t}\,ds\Big|\mathcal{F}_t\right]^{-1} \times \mathbf{E}[Z_{t,\zeta}^*|\mathcal{G}_t]\mathbf{1}_{\{t<\zeta\}} \\ &= b_\beta\mathbf{E}[Z_{t,\zeta}^*|\mathcal{G}_t]\mathbf{1}_{\{t<\zeta\}}, \end{aligned}$$

where $\frac{1}{b_\beta} = \mathbf{E}[\frac{Z_{0,\zeta}^*}{x}] = \mathbf{E}[\mathcal{Z}_{0,\zeta}^*]$. Note here that $\mathbf{E}[Z_{t,\zeta}^*|\mathcal{G}_t] = \mathbf{E}[Z_{t,\zeta^-}^*|\mathcal{G}_t]$, and thus, $Z_{t,\zeta}^* = Z_{t,\zeta^-}^*$ a.s.

The previous representation (3.7) cannot be considered as a Max-Plus decomposition of $Z$ since it only holds for $t < \zeta$. To obtain such a decomposition in this case, we must rather consider the filtration $\mathcal{G}_t$ instead of $\mathcal{F}_t$ and introduce the following process $\widetilde{Z}_t = Z_t\mathbf{1}_{\{t<\zeta\}} = Z_{t \wedge \zeta} - Z_\zeta\mathbf{1}_{\{\zeta \leq t\}}$. Thanks to the positivity of $Z$, $\widetilde{Z}$ is clearly a $\mathcal{G}_t$-supermartingale, as it is the difference between a supermartingale and a nonincreasing process. Let us also observe that $\widetilde{Z}_\zeta = 0$, which leads to the same properties as the previous case. Moreover, the positivity properties of $Z$ imply that, for $t < \zeta$,

$$Z_{t,\zeta}^* = \sup_{t \leq u \leq \zeta} Z_u = \sup_{t \leq u < \zeta} Z_u = \sup_{t \leq u < \zeta}(Z_u\mathbf{1}_{\{u<\zeta\}}) = \sup_{t \leq u < \zeta}\widetilde{Z}_u = \sup_{t \leq u \leq \zeta}\widetilde{Z}_u = \widetilde{Z}_{t,\zeta}^*.$$

Observe that, for $t = \zeta$, $\sup_{t \leq u \leq \zeta} Z_u = Z_\zeta \neq \sup_{t \leq u \leq \zeta}\widetilde{Z}_u = \widetilde{Z}_\zeta = 0$.

We hence obtain the following Max-Plus decomposition of $\widetilde{Z}$:

LEMMA 3.5. *Let* $\widetilde{Z}_t = Z_t \mathbf{1}_{\{t < \zeta\}}$ *and* $b_\beta = \frac{1}{\mathbf{E}[\widetilde{Z}^*_\zeta]}$. *Then for any* $t \leq \zeta$,

$$\widetilde{Z}_t = b_\beta \mathbf{E}[Z^*_{t,\zeta}|\mathcal{G}_t]\mathbf{1}_{\{t < \zeta\}} = b_\beta \mathbf{E}[\widetilde{Z}^*_{t,\zeta}|\mathcal{G}_t] \quad and \quad L^{\tilde{z},*}_{0,t} = b_\beta \widetilde{Z}^*_t = b_\beta Z^*_t.$$

*Then the results of Proposition 3.2 can be extended to the exponential horizon time and remain the same. We just have to replace* $Z$ *by* $\widetilde{Z}$, $b$ *by* $b_\beta$ *and the filtration* $\mathcal{F}_t$ *by* $\mathcal{G}_t$.

EXAMPLE 3.6 (Geometric Brownian motion). We show in Section 5 that if $Z$ defines a geometric Brownian motion with parameters $(-r, \sigma)$, $b_\beta = \frac{\delta - 1}{\delta}$, where $\delta$ is the root greater than $\gamma = 1 + \frac{2r}{\sigma^2}$ of the equation $y^2 - \gamma y - \frac{2\beta}{\sigma^2} = 0$.

REMARK 3.7. Equation (3.7) leads to the following representation of $Z$ on the set $\{t < \zeta\}$:

$$(3.8) \qquad Z_t = \frac{1}{b_\beta} \mathbf{E}\left[\int_t^{+\infty} \beta e^{-\beta(s-t)} Z^*_{t,s} \, ds | \mathcal{F}_t\right].$$

Let $\mu$ be a nonnegative optional random measure and let $f = f(\omega, t, x) : \Omega \times [0, +\infty] \times \mathbb{R} \to \mathbb{R}$ be a random field with the following properties:

1. For any $x \in \mathbb{R}$, the mapping $(\omega, t) \mapsto f(\omega, t, x)$ defines a progressively measurable process in $L^1(\mathbf{P}(d\omega) \otimes \mu(\omega, dt))$.

2. For any $(\omega, t) \in \Omega \times [0, +\infty]$, the mapping $x \mapsto f(\omega, t, x)$ is continuous and strictly decreasing from $+\infty$ to $-\infty$.

Then, for any given optional process $X = (X_t)_{t \in [0, +\infty]}$ with $X_{+\infty} = 0$, Bank and El Karoui have constructed in [7] a progressively measurable process $\xi = (\xi_\nu)_{\nu \in [0, +\infty)}$ with upper-right continuous paths such that

$$f\left(s, \sup_{\nu \in [t,s)} \xi_\nu\right) \mathbf{1}_{(t, +\infty]}(s) \in L^1(\mathbf{P} \otimes \mu(ds))$$

and

$$X_t = \mathbf{E}\left[\int_{(t, +\infty]} f\left(s, \sup_{\nu \in [t,s)} \xi_\nu\right) \mu(ds) | \mathcal{F}_t\right]$$

for any stopping time $t \in \mathcal{T}$. This stochastic representation of $X$ in terms of running suprema of $\xi$ comes down to equation (3.8), in the particular case where $X_t = Z_t e^{-\beta t}$, $\xi = Z$ and $f(s, l) = \beta e^{-\beta s} l$.

3.3. *Additive Lévy processes.* Let $Z$ define a supermartingale with independent additive increments and initial value $x$, such that $\mathbf{E}[Z^*_{0,\infty}] < +\infty$. Let $\mathcal{Z}$ be the process $Z$ starting from 0: $\mathcal{Z}_t = Z_t - x$. We assume the maturity $\zeta$ to be infinite.

Using the same notation as before and setting $b = \mathbf{E}[\sup_{0 \le u} Z_u] - x = \mathbf{E}[Z_{0,\infty}^*] - x = \mathbf{E}[\mathcal{Z}_{0,\infty}^*]$, it can be easily seen that $Z_t$ takes the simple form of $Z_t = \mathbf{E}[Z_{t,\infty}^* - b|\mathcal{F}_t]$, thanks to the independence property of the increments of $Z$. Then we need only to compute the expectation $\mathbf{E}[\mathcal{Z}_{0,\infty}^*]$ and apply Theorem 3.1 to solve the American Call problem. The value function is still useless in this case.

PROPOSITION 3.8.   *Let $(\mathcal{F}_t)$ be a quasi-left-continuous filtration, and $Z$ an additive Lévy process defining a supermartingale with initial value $x$, and such that $\mathbf{E}[Z_{0,\infty}^*] < +\infty$. Then:*

1. $Z_t = \mathbf{E}[Z_{t,\infty}^* - b|\mathcal{F}_t]$   *and*   $L_{0,t}^{Z,*} = Z_t^* - b$,      *where $b = \mathbf{E}[\mathcal{Z}_{0,\infty}^*]$.*
2. *An optimal stopping time to the perpetual American Call option $C_S^{\mathrm{Am}}(Z,m)$, where $S$ is a stopping time, is explicitly characterized by*

    $$T_S^Z(m) = \inf\{t \ge S; Z_t - b \ge m\} = \inf\{t \ge S; Z_t \ge m + \mathbf{E}[\mathcal{Z}_{0,\infty}^*]\},$$

    *and the exercise boundary is hence given by $E^{\mathrm{c}}(m) = m + \mathbf{E}[\mathcal{Z}_{0,\infty}^*]$.*
3. *The price at time $t$ of the perpetual American Call $C^{\mathrm{Am}}(Z,m)$ is given by*

    $$C_t^{\mathrm{Am}}(Z,m) = \mathbf{E}[(Z_{t,\infty}^* - \mathbf{E}[\mathcal{Z}_{0,\infty}^*] - m)^+|\mathcal{F}_t].$$

EXAMPLE 3.9 (Brownian motion).   Let us consider a Brownian motion with negative drift, defining a supermartingale $Z$:

$$(3.9) \qquad dZ_t = -\mu\,dt + \sigma\,dW_t, Z_0 = 0, \qquad \mu \ge 0,$$

and set $\gamma = \frac{2\mu}{\sigma^2}$. The law of $Z_{0,\infty}^*$ can be deduced from the example of the Geometric Brownian motion and we get $b = \frac{1}{\gamma}$. We can also derive a closed formula of $C_t^{\mathrm{Am}}(Z,m)$ as a function of $Z_t$.

**4. Existence of the supermartingale decomposition in the Max-Plus algebra.**   In the present section our purpose is to establish the existence of the preceding decomposition in the Max-Plus algebra. To do so, convex duality methods can be used with great efficiency. In our context, it was Whittle (see [45]) who first introduced a convex family of optimal stopping problems, in order to solve the Bandit problem. Since then, the same idea has been exploited by other authors, El Karoui–Karatzas for the Bandit problem [19], or by Bank–El Karoui [7] or El Karoui–Föllmer [15] to obtain nonlinear representation of general processes (see Remark 3.7 and Section 4.5).

4.1. *Review of the main results on optimal stopping problems.* In general, each process $X$ that we consider in the paper is defined on $[0, \zeta]$ and assumed to be adapted and càdlàg. As usual, a stopping time can take infinite values. To avoid any confusion with the horizon $\zeta$, we use the notation $\infty^+$ when it is infinite. By convention, all processes are null at this time. The family of all stopping times (finite or not) is denoted by $\mathcal{T}$ and for any stopping time $S$, $\mathcal{T}_S$ is the set of stopping times posterior to $S$: $\mathcal{T}_S = \{\tau \geq S; \tau \in \mathcal{T}\}$.

The adapted processes such that the family $\{X_{\tau \wedge \zeta}\}_{\tau \in \mathcal{T}}$ is uniformly integrable are of class $(\mathcal{D})$. [A useful criterion to show uniform integrability is the La Vallée–Poussin criterion (see [23] for further details). It states that if there exists a positive, nondecreasing, convex function $\phi(t)$ defined on $[0, \infty)$ such that $\lim_{t \to \infty} \frac{\phi(t)}{t} = +\infty$ and $\sup_{\tau \in \mathcal{T}} \mathbf{E}(\phi \circ |X_\tau|) < \infty$, then $X$ is uniformly integrable.] This property is in fact a necessary and sufficient condition for a process $X$ to be of class $(\mathcal{D})$, and it is then equivalent to the characterization given in Section 2 (see [31]).

As we shall essentially work with stopping times, the following criteria are very useful to show the path regularity ([13], Theorem 48–49, page 120):

- An optional process $X$ of class $(\mathcal{D})$ is right-continuous $(X = X^+)$ if and only if $\mathbf{E}[X_{\tau_n}\mathbf{1}_{\tau_n < \infty^+}] \to \mathbf{E}[X_\tau \mathbf{1}_{\tau < \infty^+}]$, for every *nonincreasing* sequence of stopping times $(\tau_n)_{n \geq 0} \subseteq \mathcal{T}$, with $\tau = \lim_{n \to \infty} \tau_n$ a.s.
- For the left-hand regularity, we need to be more precise. A process $X$ of class $(\mathcal{D})$ is said to be quasi-left-continuous if $\mathbf{E}[X_{\tau_n}\mathbf{1}_{0 < \tau_n < \infty^+}]$ tends to $\mathbf{E}[X_\tau \mathbf{1}_{0 < \tau < \infty^+}]$, for every *increasing* sequence of stopping times $(\tau_n)_{n \geq 0}$, tending toward $\tau$ almost surely. $\tau$ is said to be predictable.

  Then, $X$ is left-limited (with process of left limits $X^-$) and $\mathbf{E}[X_\tau^- \mathbf{1}_{\tau < \infty^+}] = \mathbf{E}[X_\tau \mathbf{1}_{\tau < \infty^+}]$ for any predictable stopping time $\tau$. Despite this equality, in general, $X^- \neq X$ and, hence, $X$ is not left-continuous. The main reason is that $X^-$ is predictable and $X$ optional. The difficulty disappears by introducing the predictable projection $X^{(p)}$ of $X$, characterized by the identity $X_\tau^{(p)} \mathbf{1}_{\{\tau < \infty^+\}} = \mathbf{E}(X_\tau | \mathcal{F}_{\tau^-}) \mathbf{1}_{\{\tau < \infty^+\}}$, for any predictable stopping time $\tau$. So $\mathbf{E}[X_{\tau^-} \mathbf{1}_{\tau < \infty^+}] = \mathbf{E}[X_\tau^{(p)} \mathbf{1}_{\tau < \infty^+}] = \mathbf{E}[X_\tau \mathbf{1}_{\tau < \infty^+}]$, and both predictable processes $X^-$ and $X^{(p)}$ are indistinguishable (cf. [13] by Dellacherie–Meyer).

  In what follows, a quasi-left-continuous process $X$ belonging to the class $\mathcal{X}_\mathcal{D}$ [of càdlàg processes of class $(\mathcal{D})$] is said to be $(\mathcal{D})$-*regular*.
- All martingales of class $(\mathcal{D})$ are regular.
- Any predictable nondecreasing process $A$ associated with the Doob–Meyer decomposition of a regular supermartingale $Z$ is continuous since it satisfies the following identity, $\mathbf{E}[A_\tau \mathbf{1}_{\tau < \infty^+}] = \mathbf{E}[A_{\tau^-} \mathbf{1}_{\tau < \infty^+}]$ for any predictable stopping time $\tau$.

Let us state a lemma, which will be useful in the sequel, on the convex transformation of quasi-left-continuous processes.

LEMMA 4.1. *If $X^- \equiv X^{(p)}$, then, for any continuous convex function $\varphi$, we have that $\varphi(X^-) \le \varphi(X)^{(p)}$. In particular, for any real $m$, $X^- \vee m \le (X \vee m)^{(p)}$.*

PROOF. For any predictable stopping time $\tau$, we have that

$$\mathbf{E}[\varphi(X_{\tau^-})\mathbf{1}_{\tau < \infty^+}] = \mathbf{E}[\varphi(X_\tau^{(p)})\mathbf{1}_{\tau < \infty^+}] = \mathbf{E}[\varphi(\mathbf{E}[X_\tau|\mathcal{F}_{\tau^-}])\mathbf{1}_{\tau < \infty^+}]$$

$$\le \mathbf{E}[\mathbf{E}[\varphi(X_\tau)|\mathcal{F}_{\tau^-}]\mathbf{1}_{\tau < \infty^+}] = \mathbf{E}[\varphi(X_\tau)^{(p)}\mathbf{1}_{\tau < \infty^+}],$$

where the inequality follows from the convex property of $\varphi$. □

4.2. *A convex family of supermartingales.* Let $Z = (Z_t; \ t \in [0, \zeta])$ be a $(\mathcal{D})$-regular supermartingale. We hence introduce the Snell envelope $Z.(m) = (Z_t(m); t \in [0, \zeta])$ of $Z \vee m = (Z_t \vee m; t \in [0, \zeta])$, that is, the smallest càdlàg supermartingale dominating $Z \vee m$, indexed by the real parameter $m$, and study the properties of this supermartingale family as a function of the parameter $m$. The key property is the convexity w.r. to m of this random field.

The dual characterization of $Z.(m)$ is well known: for any stopping time $S \in \mathcal{T}_{0,\zeta}$, $Z_S(m)$ is in the form of

$$(4.1) \qquad Z_S(m) = \operatorname*{ess\,sup}_{\tau \in \mathcal{T}_{S,\zeta}} \mathbf{E}[Z_\tau \vee m|\mathcal{F}_S],$$

where $\mathcal{T}_{S,\zeta}$ denotes the collection of $\mathcal{F}$-stopping times with values in $[S, \zeta]$. The following observations are useful:

- If $Z$ is a martingale $M$, then $M \vee m$ is a submartingale and it is never optimal to stop before maturity. Thus, $M.(m)$ is a martingale, and

$$(4.2) \qquad M_S(m) = \mathbf{E}[M_\zeta \vee m|\mathcal{F}_S].$$

- When $Z$ is a $(\mathcal{D})$-regular supermartingale and thus of the class $(\mathcal{D})$, there exist two u.i. martingales $U$ and $V$, such that $U \le Z \le V$; then the Snell envelope $Z.(m)$ belongs to the class $(\mathcal{D})$, and $\mathbf{E}[U_\zeta \vee m|\mathcal{F}_S] \le Z_S(m) \le \mathbf{E}[V_\zeta \vee m|\mathcal{F}_S]$.

The main properties of $(Z_S(m))_{S \in \mathcal{T}_{0,\zeta}}$ essentially lie on properties related to the Snell envelope and optimal stopping theory. The following proposition precises the regularities of this parameterized family of supermartingales with respect to the parameter $m$.

PROPOSITION 4.2. *1. There exists a 1-Lipschitzian regular version of the random field $(S, m) \mapsto Z_S(m)$, such that $m \mapsto Z_S(m)$ is convex and nondecreasing and $m \mapsto Z_S(m) - m$ is convex, nonincreasing. Moreover, for any $\mathcal{F}_S$-measurable r.v. $\Lambda_S$,*

$$Z_S(\Lambda_S) = \operatorname*{ess\,sup}_{S \le \tau \le \zeta} \mathbf{E}[Z_\tau \vee \Lambda_S|\mathcal{F}_S], \qquad \tau \in \mathcal{T}_{S,\zeta}.$$

*2. The boundary $m$ is absorbing for the process $Z.(m)$. In particular, if*

$$\theta_S(m) := \inf\{t \in [S, \zeta]; Z_t(m) = m\}, \qquad (= \infty^+ \text{ if } \{\cdot\} = \varnothing),$$

*then, on the set $\{\theta_t(m) \le \zeta\}$, $Z_U(m) = m$ for all $U \in [\theta_t(m), \zeta]$.*

PROOF.   1. The proof of the first item is technical and relegated to the Appendix.

2. It is a classic property of nonnegative supermartingales that 0 is absorbing. Applying this result to the nonnegative supermartingale $Z.(m) - m$, we obtain the desired result. The same property holds when $m$ is replaced by an $\mathcal{F}_S$-measurable r.v. $\Lambda_S$.   □

4.2.1. *Optimal stopping times.*   In this section we briefly outline some basic facts concerning the optimal stopping theory without proof (see [14] and [32] for more general treatments), and especially the links between the Snell envelope and optimal stopping times. We directly express these results in terms of the process $Z.(m)$. The filtration $(\mathcal{F}_t)$ is assumed to be quasi-left-continuous in the following theorem.

THEOREM 4.3.   *Assume $Z$ to be a $(\mathcal{D})$-regular supermartingale. Then:*
*1. $Z.(m)$ is a $(\mathcal{D})$-regular supermartingale.*
*2. Moreover, let $T_S(m) := \inf\{t \in [S, \zeta]; Z_t(m) = Z_t\}, = \infty^+$ if the set is empty.*
*The stopping time $T_S(m) \wedge \zeta$ is an optimal stopping time:*

$$Z_S(m) = \mathbf{E}[\sup(Z_{T_S(m) \wedge \zeta}, m) | \mathcal{F}_S] \quad \text{and} \quad Z_{T_S(m) \wedge \zeta}(m) = \sup(Z_{T_S(m) \wedge \zeta}, m),$$

*but the smallest optimal stopping time is $\widehat{T}_S(m) := T_S(m) \wedge \theta_S(m)$.*
*3. The family $m \mapsto T_S(m) \wedge \zeta$ is nondecreasing and left-continuous.*

REMARK 4.4.   *The proof of the left-continuity of the family $m \mapsto T_S(m) \wedge \zeta$ is the most technical part of the work. It can be omitted in a first reading.*

PROOF OF THEOREM 4.3.   1. If the supermartingale $Z$ is continuous (in the case of a Brownian filtration, e.g.), the process $Z \vee m$ is also continuous, and the classic theory may be applied to explain the first part of the theorem (see [19]). In the general case where $Z$ is only quasi-left-continuous, the process $Z \vee m$ is only upper-quasi-left-continuous, since from Lemma 4.1, $Z^- \vee m \le (Z \vee m)^{(p)}$. Equivalently and in terms of stopping times, for any increasing sequence of stopping times $S_n \uparrow S$, $\lim_{n \to \infty} \mathbf{E}[Z_{S_n} \vee m] \le \mathbf{E}[Z_S \vee m]$ (cf. the second preliminary remark). To conclude, we have to use general results on optimal stopping problems; in particular, the desired result follows immediately from [14], Theorem 2.43, page 142.

2. We first observe that

$$Z_{T_S(m) \wedge \zeta}(m) = Z_{T_S(m)}(m) \mathbf{1}_{\{T_S(m) \leq \zeta\}} + \sup(Z_\zeta, m) \mathbf{1}_{\{T_S(m) = \infty^+\}}$$
$$= Z_{T_S(m)} \mathbf{1}_{\{T_S(m) \leq \zeta\}} + \sup(Z_\zeta, m) \mathbf{1}_{\{T_S(m) = \infty^+\}}.$$

But on $\{T_S(m) \leq \zeta\}$, $Z_{T_S(m)}(m) = Z_{T_S(m)} \geq m$ and so

$$Z_{T_S(m) \wedge \zeta}(m) = \sup(Z_{T_S(m) \wedge \zeta}, m).$$

It remains to observe that

$$Z_{\widehat{T}_S(m)}(m) = Z_{\theta_S(m) \wedge (T_S(m) \wedge \zeta)}(m)$$

$$= \sup(Z_{T_S(m) \wedge \zeta}, m) \mathbf{1}_{\{T_S(m) \wedge \zeta \leq \theta_S(m)\}} + m \mathbf{1}_{\{\theta_S(m) < T_S(m) \wedge \zeta\}}.$$

Then since $m$ is absorbing on the set $\{\theta_S(m) < T_S(m) \wedge \zeta\}$, $Z_{T_S(m) \wedge \zeta}(m) = m = \sup(Z_{T_S(m) \wedge \zeta}, m)$. So we finally obtain that $Z_{\widehat{T}_S(m)}(m) = Z_{T_S(m) \wedge \zeta}(m) = \sup(Z_{T_S(m) \wedge \zeta}, m)$.

3. Let $\epsilon > 0$. Since $m \mapsto Z_U(m)$ is Lipschitz continuous at any time $U \geq S$, $Z_U(m - \epsilon) \to Z_U(m)$ a.s. and in $L^1$, as $\epsilon \downarrow 0$.

Let us set $T_S^\epsilon := T_S(m - \epsilon)$, $T_S := T_S(m)$ and $T_{S-} := \lim \uparrow T_S(m - \epsilon)$. We specify the predictable part $T_{S-}^{(p)}$ of $T_{S-}$ as the stopping time defined by

$$T_{S-}^{(p)} = T_{S-} \quad \text{on } H_{T_S}^- = \{\omega, T_S^\epsilon \uparrow\uparrow T_{S-}\}, \qquad = \infty^+ \text{ if not,}$$

$$T_{S-}^{(s)} = T_{S-} \quad \text{on } (H_{T_S}^-)^c = \{\omega, \exists \epsilon \; T_S^\epsilon = T_{S-}\}, \qquad = \infty^+ \text{ if not.}$$

So $T_{S-} = \inf(T_{S-}^{(p)}, T_{S-}^{(s)})$. We can give a precise description of the limit of the different terms:

(a) on $(H_{T_S}^-)^c$, $\lim Z_{T_S^\epsilon \wedge \zeta}(m - \epsilon) = Z_{T_{S-} \wedge \zeta}(m)$, and if $T_{S-} \leq \zeta$, then $T_S^\epsilon \leq \zeta$ and $Z_{T_S^\epsilon \wedge \zeta} = Z_{T_S^\epsilon}$ tends to $Z_{T_{S-}}$. If $T_{S-} = \infty^+$, then there exists $\epsilon$ such that $T_S^\epsilon = \infty^+$ and $Z_{T_S^\epsilon \wedge \zeta}(m - \epsilon) = \sup(Z_\zeta, m - \epsilon)$. So in both cases, $Z_{T_{S-} \wedge \zeta}(m) = \sup(Z_{T_{S-} \wedge \zeta}, m)$. We easily deduce that $T_{S-}(m) = T_S(m)$ on $(H_{T_S}^-)^c$.

(b) on $H_{T_S}^-$, thanks to the strict monotonicity of the sequence $T_S^\epsilon$, $T_{S-}^{(p)} \leq \zeta$, and $\lim_{\epsilon \to 0} Z_{T_S^\epsilon \wedge \zeta}(m - \epsilon) = Z_{T_{S-}^{(p)}}^-(m)$. On the other hand, $Z_{T_S^\epsilon}(m - \epsilon) = Z_{T_S^\epsilon}$ since $T_S^\epsilon \leq \zeta$ a.s., whence $\lim_{\epsilon \to 0} Z_{T_S^\epsilon \wedge \zeta}(m - \epsilon) = \lim_{\epsilon \to 0} Z_{T_S^\epsilon} = Z_{T_{S-}^{(p)}}^-$.

Since $T_{S-}^{(p)}$ is predictable on $H_{T_S}^-$, and thanks to the quasi-left-continuity of $Z$ and $Z.(m)$, this yields

$$Z_{T_{S-}^{(p)}}^-(m) = Z_{T_{S-}^{(p)}}^{(p)}(m) = Z_{T_{S-}^{(p)}}^- = Z_{T_{S-}^{(p)}}^{(p)} = \mathbf{E}[Z_{T_{S-}^{(p)}} | \mathcal{F}_{T_{S-}^{(p)}}^-].$$

Then on $H_{T_S}^-$, $Z_{T_{S-}^{(p)}}(m) = Z_{T_{S-}^{(p)}}$ a.s. and $T_{S-}^{(p)} = T_{S-} = T_S(m)$, since $T_S(m)$ is the smallest stopping time $U$ after $S$ satisfying $Z_U(m) = Z_U$.  $\square$

REMARK 4.5.   If the supermartingale $Z$ is not quasi-left-continuous, it has been shown in [14] that we can find an optimality loss in the optimal stopping problem of the process $Z \vee m$; more precisely, the candidate to be an optimal stopping time is the time $\hat{\tau}_S(m)$ such that

$$\hat{\tau}_S(m) := \inf\{t \geq S;\ Z_t(m) = Z_t \vee m \text{ or } Z_t^-(m) = Z_t^- \vee m\} \wedge \zeta.$$

It is possible to adapt the previous proof by introducing the two families of stopping times $\widetilde{\theta}_S(m) = \inf\{t \geq S; Z_t(m) = m \text{ or } Z_t^-(m) = m\}$ and $\widetilde{T}_S(m) = \inf\{t \geq S; Z_t(m) = Z_t \text{ or } Z_t^-(m) = Z_t^-\}$.

4.3. *Convex analysis and characterization of the $Z$-Max-Plus decomposition.*   In what follows, we assume $Z$ to be a $(\mathcal{D})$-regular supermartingale. We are still working with the 1-Lipschitzian version of the random field $m \mapsto Z_.(m)$. Thanks to the convexity and monotonicity of $m \mapsto Z_. \vee m$, the random field $Z_.(m)$ inherits the same properties. In particular, it is possible to characterize the left-hand derivative of $Z_S(m)$ w.r.t. $m$.

Similar ideas can be found in the paper by El Karoui and Karatzas [19]. By considering some properties (w.r.t. $K$) of a convex family of American Put options, they have provided a representation of the early exercise premium of an American put-option with given strike-price $K > 0$, on a finite time-horizon. The main difference in our context is that the asset process is a supermartingale which is not required to be continuous, but only continuous in expectation with respect to stopping times. Also very closed to this point of view is the representation theorem of any process $X$ of class $(\mathcal{D})$ established by Bank and El Karoui [7]. We assume in the following proposition that the filtration $(\mathcal{F}_t)$ is quasi-left-continuous.

PROPOSITION 4.6 [Static representation of $Z_t(m)$].   *Fix $S$ in $\mathcal{T}_{0,\zeta}$ and let $\Lambda_S(\alpha)$ be the left-inverse at time $\alpha$ of the mapping $m \mapsto T_S(m)$, where $m \in \mathbb{R}$: $\Lambda_S(\alpha) := \sup\{m; T_S(m) \leq \alpha\}$, with the conventional notation $\sup\{\varnothing\} = -\infty$.*

(a) *The convex mapping $m \mapsto Z_S(m)$ has a left-hand derivative satisfying*

$$
\begin{aligned}
(4.3) \qquad \frac{\partial^-}{\partial m} Z_S(m) &= \mathbf{P}[Z_\zeta < m; T_S(m) = \infty^+ | \mathcal{F}_S] \\
&= \mathbf{P}[Z_\zeta < m; \Lambda_S(\zeta) < m | \mathcal{F}_S] = 1 + \frac{\partial^-}{\partial m} C_S^{\mathrm{Am}}(Z, m) \qquad a.s.
\end{aligned}
$$

(b) *For all reals $m$,*

$$Z_S(m) = \mathbf{E}[\Lambda_S(\zeta) \vee Z_\zeta \vee m | \mathcal{F}_S] \quad and \quad Z_S = Z_S(-\infty) = \mathbf{E}[\Lambda_S(\zeta) \vee Z_\zeta | \mathcal{F}_S].$$

This result is strongly related to "traditional envelope theorems" which describe sufficient conditions for the value of a parameterized optimization problem to be differentiable in the parameter and provide a formula for the derivative. See [37] for a review of envelope theorems.

PROOF OF PROPOSITION 4.6. (a) The proof is based on the optimality of $T_S(m) \wedge \zeta$ and on the classic convex inequalities

$$\epsilon \mathbf{1}_{\{x < m - \epsilon\}} \leq x \vee m - x \vee (m - \epsilon) \leq \epsilon \mathbf{1}_{\{x < m\}}.$$

The set $\{Z_{T_S(m) \wedge \zeta} < m\}$ plays a key role. Since on $\{T_S(m) \leq \zeta\}$, $Z_{T_S(m)} = Z_{T_S(m)}(m) \geq m$,

$$\{Z_{T_S(m) \wedge \zeta} < m\} = \{Z_\zeta < m; T_S(m) = \infty^+\}.$$

These observations yield to the following series of inequalities:

$$Z_S(m) - Z_S(m - \epsilon) \leq \mathbf{E}[\sup(Z_{T_S(m) \wedge \zeta}, m) - \sup(Z_{T_S(m) \wedge \zeta}, m - \epsilon) | \mathcal{F}_S]$$
$$\leq \epsilon \mathbf{P}[Z_\zeta < m; T_S(m) = \infty^+ | \mathcal{F}_S].$$

Now to obtain a lower bound, we apply the optimality to $T_S(m - \epsilon) \wedge \zeta$:

$$Z_S(m) - Z_S(m - \epsilon) \geq \mathbf{E}[\sup(Z_{T_S(m-\epsilon) \wedge \zeta}, m) - \sup(Z_{T_S(m-\epsilon) \wedge \zeta}, m - \epsilon) | \mathcal{F}_S]$$
$$\geq \epsilon \mathbf{P}[Z_\zeta < m - \epsilon; T_S(m - \epsilon) = \infty^+ | \mathcal{F}_S].$$

Since the sequence $T_S(m)$ is nondecreasing and left-continuous,

$$(4.4) \quad \begin{aligned} \lim_{\epsilon \downarrow 0} \mathbf{P}[Z_\zeta < m - \epsilon; T_S(m - \epsilon) = \infty^+ | \mathcal{F}_S] \\ \leq \mathbf{P}[Z_\zeta < m; T_S(m) = \infty^+ | \mathcal{F}_S]. \end{aligned}$$

However, $\{T_S(m) = \infty^+\} = \bigcap_\epsilon \{T_S(m - \epsilon) = \infty^+\}$ since, on $H_{T_S}^-$, $T_S(m) = \lim T_S(m - \epsilon) \leq \zeta$. The inequality in (4.4) is therefore an equality and we finally obtain

$$\frac{\partial^-}{\partial m} Z_S(m) = \lim_{\epsilon \downarrow 0} \mathbf{P}[Z_\zeta < m - \epsilon; T_S(m - \epsilon) = \infty^+ | \mathcal{F}_S]$$
$$= \mathbf{P}[Z_\zeta < m; T_S(m) = \infty^+ | \mathcal{F}_S].$$

(b) Now, we would like to reintegrate in order to derive an explicit representation of $Z_S(m)$. To do so, we need to express the event $\{T_S(m) = \infty^+\}$ according to $m$ in a simpler way. Note that $\{m; T_S(m) \leq \zeta\}$ is an interval closed on the right by $\Lambda_S(\zeta)$, also defined by

$$\Lambda_S(\zeta) = \sup\{m; T_S(m) \leq \zeta\}, \qquad (= -\infty \text{ if } \{\cdot\} = \varnothing).$$

It follows that $\frac{\partial^-}{\partial m} Z_S(m) = \mathbf{P}[Z_\zeta < m; m > \Lambda_S(\zeta) | \mathcal{F}_S]$. Then we need some boundary conditions. Since $Z_S(m) - m = \operatorname{ess\,sup}_{\tau \in \mathcal{T}_{S, \zeta}} \mathbf{E}[(Z_\tau - m)^+ | \mathcal{F}_S]$, we can use (4.2) to show that if $Z$ is dominated by the uniformly integrable martingale $V$,

$$0 \leq \lim_{m \uparrow +\infty} (Z_S(m) - m) \leq \lim_{m \uparrow +\infty} \mathbf{E}[(V_\zeta - m)^+ | \mathcal{F}_S] = 0$$

and we can hence write

$$Z_S(m) - m = \int_m^{+\infty} -\frac{\partial^-}{\partial \alpha}(Z_S(\alpha) - \alpha)\, d\alpha.$$

Note that we cannot directly reintegrate $\frac{\partial^-}{\partial m} Z_S(m)$, since the limit of $Z_S(m)$ when $m$ goes to $-\infty$ is equal to $Z_S$ and thus unknown. Then, the following equality can be deduced from (4.3):

$$Z_S(m) - m = \int_m^{+\infty} \mathbf{P}[\Lambda_S(\zeta) \vee Z_\zeta \geq \alpha | \mathcal{F}_S]\, d\alpha.$$

Applying the conditional Fubini theorem, we hence obtain

$$Z_S(m) - m = \mathbf{E}[(\Lambda_S(\zeta) \vee Z_\zeta - m)^+ | \mathcal{F}_S]$$

and

$$Z_S(m) = \mathbf{E}[\Lambda_S(\zeta) \vee Z_\zeta \vee m | \mathcal{F}_S].$$

(c) Now letting $m \downarrow -\infty$ and applying the monotonous convergence theorem, we get $\lim_{m \downarrow -\infty} Z_S(m) = Z_S(-\infty) = \mathbf{E}[\Lambda_S(\zeta) \vee Z_\zeta | \mathcal{F}_S]$. On the other hand, the sequence $T_S(m)$ is decreasing to $T_S^+(-\infty)$ and by the right-continuity of $Z$, $Z_{T_S(m) \wedge \zeta} \vee m$ goes to $Z_{T_S^+(-\infty) \wedge \zeta}$. By the Lebesgue theorem,

$$Z_S \leq \lim_{m \downarrow -\infty} Z_S(m) = \lim_{m \downarrow -\infty} \mathbf{E}[\sup(Z_{T_S(m) \wedge \zeta}, m) | \mathcal{F}_S]$$

$$= \mathbf{E}[Z_{T_S(-\infty)^+ \wedge \zeta} | \mathcal{F}_S] \leq Z_S, \qquad \text{a.s.,}$$

since $T_S(-\infty)^+ \wedge \zeta$ is a stopping time posterior to $S$ and $Z$ a supermartingale. It finally follows that $Z_S = Z_S(-\infty) = \mathbf{E}[\Lambda_S(\zeta) \vee Z_\zeta | \mathcal{F}_S]$. $\quad\square$

The preceding proposition provides a static representation of $Z_t(m)$ ($t$ fixed). We shall hereafter exploit the dynamic structure of the Snell envelope $\{Z_t(m); t \geq 0\}$ in order to deduce a representation which exhibits the dependence of $\Lambda_t(\zeta)$ w.r.t. $t$. To do that, we use techniques related to change of variable in $\mathbb{R}$, that we recall in the following lemma:

LEMMA 4.7.   *Set*

$$\Lambda_t(\alpha) := \sup\{m; T_t(m) \leq \alpha\}; \qquad \alpha \in [t, \zeta], \ = -\infty \ \text{if the set is empty.}$$

*Then $\Lambda_t(\alpha)$ defines the right-continuous inverse of the nondecreasing, left-continuous mapping $m \mapsto T_t(m)$. In other terms,*

$$T_t(m) \leq \alpha \quad \Longleftrightarrow \quad m \leq \Lambda_t(\alpha) \qquad \forall \alpha \in [t, \zeta].$$

Note that since the inequalities are large, $\Lambda_t(\alpha)$ represents the right-hand inverse of the left-continuous process $m \mapsto T_t(m)$. Previous works on time changes [22] have shown that the process $\alpha \mapsto \Lambda_t(\alpha)$ is right-continuous. Let us stress that we are not in the usual cases, where we rather consider right-hand inverses of right-continuous processes, and where the inequalities are consequently strict. The interested reader can, for example, refer to [22] for full details on time change theory.

Now, the following theorem represents the increasing process $\{\Lambda_t(\alpha); t \leq \alpha \leq \zeta\}$ in terms of a running supremum process and gives an explicit form of the martingale $M^\oplus$ of the Max-Plus decomposition of $Z$.

THEOREM 4.8.   *Assume that the filtration $(\mathcal{F}_t)$ is quasi-left-continuous, and let $L_t$ be the $\mathcal{F}_t$-measurable r.v. defined by*

$$L_t := \sup\{m \in \mathbb{Q}; Z_t(m) = Z_t\}$$
$$= \sup\{m \in \mathbb{Q}; T_t(m) = t\}, \qquad = -\infty \ \text{if the set is empty.}$$

*Let $L_{t,\alpha}^*$ be the running supremum of $L$ over $[t, \alpha]$ ($t \leq \alpha \leq \zeta$), that is, $L_{t,\alpha}^* = \sup_{t \leq s \leq \alpha} L_s$. Then:*

(a) *$L_{t,\alpha}^* = \Lambda_t(\alpha)$ for any $\alpha \in [t, \zeta]$ and, hence, $Z_t(m) = \mathbf{E}[L_{t,\zeta}^* \vee Z_\zeta \vee m | \mathcal{F}_t]$, and $\alpha \mapsto L_{t,\alpha}^*$ is right-continuous for any $\alpha \in [t, \zeta[$.*

(b) *The process $(M_t)_{t \geq 0}$ defined by*

$$M_t = \mathbf{E}[L_{0,\zeta}^* \vee Z_\zeta | \mathcal{F}_t] = Z_t(L_{0,t}^*) \geq Z_t = Z_t(-\infty), \qquad 0 \leq t \leq \zeta,$$

*is the martingale $M^\oplus$ of the Max-Plus decomposition of $Z$ since the increasing process $(L_{0,t}^*)$ satisfies the flat-off condition*

$$\int_{[0,\zeta]} (M_t - Z_t) \, dL_{0,t}^* = 0 \qquad a.s.$$

PROOF.   (a) First, note that the increase property and the continuity of the mapping $m \mapsto Z_t(m)$ imply that $L_t$ is the right-point of the closed interval $\{m, Z_t(m) = Z_t\}$.

Then note that, for $\alpha \leq \zeta$, $T_t(m) \leq \alpha$, if and only if, there exists $s \in [t, \alpha]$ s.t. $Z_s(m) = Z_s$, or equivalently, $L_s \geq m$. This consequently leads to the following series of identities:

$$\{T_t(m) = t\} = \bigcap_{\alpha > 0} \{T_t(m) \leq t + \alpha\}$$

$$= \bigcap_{\alpha > 0} \{\exists s \in [t, t+\alpha]; L_s \geq m\} = \left\{\limsup_{s \downarrow t} L_s \geq m\right\}.$$

However, since $T_t(m) = t$ if and only if $L_t \geq m$, it immediately follows that $L_t = \limsup_{s \downarrow t} L_s$ and $L_t$ is upper-right semi-continuous. We hence get

$$T_t(m) \leq \alpha \quad \Longleftrightarrow \quad \exists s \in [t, \alpha]; \qquad m \leq L_s \quad \Longleftrightarrow \quad m \leq L_{t,\alpha}^*.$$

This last equivalence in conjunction with the preceding lemma easily implies that $L_{t,\alpha}^* = \Lambda_t(\alpha)$ for any $\alpha \in [t, \zeta]$.

(b) $(L_{0,t}^*)_{t \geq 0}$ is obviously an increasing process. Let $S$ be a stopping time corresponding to an increasing point of $L_{0,\cdot}^*$.

If $S < \zeta$, it necessarily satisfies $\sup_{0 \leq t \leq \zeta} L_t = \sup_{S \leq t \leq \zeta} L_t$, whence $M_S = Z_S$.

If $S = \zeta$, we have that $L_{0,S^-}^* < L_S$, otherwise the increasing process does not jump at $\zeta$. However, $L_\zeta = Z_\zeta$ and $M_\zeta = L_{0,\zeta^-}^* \vee L_\zeta$. It immediately follows that $M_\zeta = Z_\zeta$ and the flat-off condition is well satisfied. Thus, by Theorem 2.8, this immediately yields the uniqueness of the martingale in the Max-Plus decomposition $M_S = Z_S \oplus L_{0,S}^*$, where $S$ satisfies $L_S = L_{0,S}^*$. $\square$

Proposition 4.6 together with Theorem 4.8 allows to study the regularity of the derivative of the value function $C_S^{\mathrm{Am}}(Z, m)$ w.r.t. the strike $m$. We say that there is smooth pasting if the two derivatives coincide at the optimal stopping boundary determined by $m = L_S = \sup\{m; C_S^{\mathrm{Am}}(Z, m) = Z_S - m\}$. The following corollary gives a full description of the right-hand derivative jumps at the boundary and establishes the conditions under which smooth pasting occurs.

COROLLARY 4.9 (Smooth-fit principle). 1. *The left-hand and right-hand derivatives of the American Call price* $C_S^{\mathrm{Am}}(Z, m)$ *w.r.t.* $m$ *can be expressed as*

$$\frac{\partial^-}{\partial m} C_S^{\mathrm{Am}}(Z, m) = -\mathbf{P}[Z_\zeta \vee L_{S,\zeta}^* \geq m | \mathcal{F}_S],$$

$$\frac{\partial^+}{\partial m} C_S^{\mathrm{Am}}(Z, m) = -\mathbf{P}[Z_\zeta \vee L_{S,\zeta}^* > m | \mathcal{F}_S] \qquad a.s.$$

2. *The left-hand derivative is almost surely continuous at the boundary* $m = L_S$:

$$\frac{\partial^-}{\partial m} C_S^{\mathrm{Am}}(Z, m)\Big|_{m=L_S} = \frac{\partial^-}{\partial m}(Z_S - m)\Big|_{m=L_S} = -1.$$

3. *The right-hand derivative is not continuous at the boundary if*

$$(4.5) \quad \mathbf{P}[Z_\zeta \vee L_{S,\zeta}^* = L_S | \mathcal{F}_S] = \mathbf{P}[L_t \leq L_S, Z_\zeta \leq L_S \ \forall t \in [S, \zeta] | \mathcal{F}_S] > 0,$$

*and, hence, the smooth-fit principle does not occur if the conditional distribution function of* $Z_\zeta \vee L_{S,\zeta}^*$ *jumps at the point* $L_S$.

The first item of Corollary 4.9 follows from the convexity property of the mapping $m \mapsto Z_S(m)$, which implies that $\frac{\partial^+}{\partial m} Z_S(m) = \lim_{\epsilon \downarrow 0} \frac{\partial^-}{\partial m} Z_S(m + \epsilon)$.

It is interesting to note that, when the underlying $Z$ evolves according to a positive multiplicative Lévy process, the smooth-fit principle we have defined here is equivalent to the "classic" one usually met in the context of American Put options (see [3], Theorem 6). In fact, in light of equation (3.5), the American Call prices are transformed into "classic" American Put prices with a discount factor, by simply exchanging the strike and the spot price of the underlying $(C_0^{\mathrm{Am}}(Z, m) = \mathrm{Put}_0^{\mathrm{Am}}(m, Z_0))$, and so any point $m$ of our boundary corresponds to a point $x$ of the American Put boundary such that $L(x) = bx = m$.

REMARK 4.10. Recall that the stopping time $T_S(m) \wedge \zeta$ is optimal, whereas $T_S(m)$ is not. Moreover, the smallest optimal stopping time is $\widehat{T}_S(m) = T_S(m) \wedge \theta_S(m)$, where $\theta_S(m)$ is defined as the first stopping time after $S$ at which $Z_.(m)$ equals $m$.

The mapping $m \mapsto \theta_S(m)$ is nonincreasing, with effective domain $\mathrm{Dom}_S := \{m; \theta_S(m) \leq \zeta\} = [K_{S,\zeta}, +\infty]$, where $K_{S,\zeta}$ is the smallest value of $m$ for which $Z_S(m) - m = 0$, that is,

$$K_{S,\zeta} := \mathcal{F}_S - \mathrm{ess\,sup}\, Z_{S,\zeta}^* = \mathrm{essinf}\{Y_S | Y_S \in \mathcal{F}_S, Y_S \geq Z_{S,\zeta}^* \text{ a.s.}\}.$$

Observe the following:

• If $S$ and $T$ are two stopping times such that $S \leq T$, then $K_{S,\zeta} \geq K_{T,\zeta}$. In fact, since any $\mathcal{F}_S$-measurable variable is $\mathcal{F}_T$-measurable, we have

$$K_{S,\zeta} = \mathrm{ess\,inf}\{Y \in \mathcal{F}_S, Y \geq Z_{S,\zeta}^* \geq Z_{T,\zeta}^* \text{ a.s.}\}$$

$$\geq \mathrm{ess\,inf}\{Y \in \mathcal{F}_T, Y \geq Z_{T,\zeta}^* \text{ a.s.}\} = K_{T,\zeta}.$$

• For all stopping times $U \geq S$, $Z_S(K_{S,\zeta}) = Z_U(K_{S,\zeta}) = K_{S,\zeta}$, a.s.

4.4. *Max-Plus decomposition and American options.* We focused in Section 3 on the problem to find an optimal stopping time to an American Call option written on a supermartingale underlying of class $(\mathcal{D})$.

In the present section we are concerned with more general American Call options without discount factor, written on a $(\mathcal{D})$-regular underlying $Y$ not necessarily defining a supermartingale. We denote by $Z^Y$ the Snell envelope of $Y$, which also defines a $(\mathcal{D})$-regular process by [14, 31]. The filtration $(\mathcal{F}_t)$ is assumed to be quasi-left-continuous in the following theorem.

THEOREM 4.11. *Let $Y$ be a $(\mathcal{D})$-regular process, $Z^Y$ the Snell envelope of $Y$ and $C_.^{\mathrm{Am}}(Y, m)$ the price of an American Call option, given at any stopping time $S \leq \zeta$ by,*

$$C_S^{\mathrm{Am}}(Y, m) = \mathrm{ess\,sup}_{\tau \geq S} \mathbf{E}[(Y_\tau - m)^+ | \mathcal{F}_S].$$

1. *The American Calls written on $Y$ and $Z^Y$, respectively, have the same price:*

$$C_S^{\mathrm{Am}}(Y,m) = C_S^{\mathrm{Am}}(Z^Y,m).$$

2. *Let $L_\cdot^Z$ be an index process in the Max-Plus decomposition of $Z^Y$. The stopping time $T_S^Z(m) \wedge \zeta := \inf\{t \geq S; L_t^Z \geq m\} \wedge \zeta$ is an optimal stopping time.*

   Intrinsic characterization in terms of $Y$.

3. *Define the stopping times $D_S$ and $T_S^Y(m)$ by*

$$D_S := \inf\{t \geq S; Z_t^Y = Y_t\},$$

$$T_S^Y(m) := \inf\{t \geq S; C_t^{\mathrm{Am}}(Y,m) = Y_t - m\}.$$

   *Then $D_{T_S^Z(m) \wedge \zeta} = T_S^Y(m) \wedge \zeta$ and $m \mapsto T_S^Y(m) \wedge \zeta$ is nondecreasing and left-continuous.*

4. *Set $L_S^Y := \sup\{m; C_S^{\mathrm{Am}}(Y,m) = Y_S - m\}$ and $L_{S,u}^{Y,*} = \sup_{s \leq t \leq u} L_t^Y$. Then $L_{S,\zeta}^{Z,*} = L_{S,\zeta}^{Y,*}$ for any stopping time $S \leq \zeta$ and $C_S^{\mathrm{Am}}(Y,m)$ has a closed formula given by*

$$C_S^{\mathrm{Am}}(Y,m) = \mathbf{E}[(L_{S,\zeta}^{Z,*} \vee Z_\zeta - m)^+ | \mathcal{F}_S] = \mathbf{E}[(L_{S,\zeta}^{Y,*} \vee Y_\zeta - m)^+ | \mathcal{F}_S].$$

PROOF. 1. First note that $C_t^{\mathrm{Am}}(Z^Y,m) = \operatorname{ess\,sup}_{t \leq S \leq \zeta} \mathbf{E}[Z_S^Y \vee m | \mathcal{F}_t] - m := Z_t^Y(m) - m$.

Let $Z'$ be a supermartingale dominating $Y \vee m$. Since $Z'$ dominates $Y$, it also dominates its Snell envelope $Z^Y$. We immediately deduce that $Z'$ dominates $Z^Y \vee m$, and its Snell envelope $Z_\cdot^Y(m)$ as well. Hence, the Snell envelope of $Y \vee m$ dominates $Z_\cdot^Y(m)$, and as the reverse inequality trivially holds, the desired result follows at once.

2. At the beginning of this section, this point has been already proven for any index process $L$ satisfying the Max-Plus decomposition of $Z$. We here give another specific proof for the index process $L$ constructed in Section 4.

Thanks to Theorem 4.3, the stopping time $T_S^Z(m) \wedge \zeta$ defined by

$$T_S^Z(m) \wedge \zeta = \inf\{u \geq S; C_u^{\mathrm{Am}}(Y,m) = Z_u^Y - m\} \wedge \zeta$$

is optimal. Let $L_\cdot^Z$ be the index process in the Max-Plus decomposition of $Z^Y$:

$$L_S^Z = \sup\{m; Z_S^Y(m) = Z_S^Y\} = \sup\{m; C_S^{\mathrm{Am}}(Z^Y,m) = Z_S^Y - m\}.$$

$T_S^Z(m)$ defines the left-hand inverse of $\alpha \mapsto L_{S,\alpha}^{Z,*}$ and thus satisfies the above relation in Theorem 4.11.

3. To simplify the notation in the proof, we will omit the parameter $m$ from the expressions of both stopping times $T_S^Z(m)$ and $T_S^Y(m)$.

Set $D_S(m) := D_{T_S^Z \wedge \zeta} = \inf\{t \geq T_S^Z \wedge \zeta; Z_t^Y = Y_t\}$. Since $D_S(m)$ is the first stopping time after $T_S^Z \wedge \zeta$, at which $Y$ reaches its Snell envelope $Z^Y$, we immediately get by the optimal stopping theory

$$(4.6) \qquad Z_{T_S^Z \wedge \zeta}^Y = \mathbf{E}[Y_{D_S(m)} | \mathcal{F}_{T_S^Z \wedge \zeta}].$$

Thanks to the optimality of the stopping time $T_S^Z \wedge \zeta$ for $Z_\cdot^Y(m)$, $Z_{T_S^Z \wedge \zeta}^Y(m) = \sup(Z_{T_S^Z \wedge \zeta}^Y, m)$ and, hence,

$$
\begin{aligned}
(4.7) \qquad Z_{T_S^Z \wedge \zeta}^Y(m) &= \mathbf{E}[Y_{D_S(m)} | \mathcal{F}_{T_S^Z \wedge \zeta}] \vee m \\
&\leq \mathbf{E}[\sup(Y_{D_S(m)}, m) | \mathcal{F}_{T_S^Z \wedge \zeta}] \\
&\leq Z_{T_S^Z \wedge \zeta}^Y(m).
\end{aligned}
$$

Thus, the chain of inequalities is indeed a series of equalities which proves that $D_S(m)$ is an optimal stopping time starting from $T_S^Z \wedge \zeta$.

Now note that $Z_{T_S^Z}(m) = Z_{T_S^Z} \geq m$ on the set $\{T_S^Z \leq \zeta\}$. This observation together with (4.6) implies that $Y_{D_S(m)} \geq m$ on $\{T_S^Z \leq \zeta\}$. This allows to reinterpret $D_S(m)$ as $T_S^Y$. In fact, on the set $\{T_S^Z \leq \zeta\}$,

$$D_S(m) = \inf\{t \geq T_S^Z; Z_t^Y(m) = Z_t^Y = Y_t\} = \inf\{t \geq S; Z_t^Y = Z_t^Y = Y_t\},$$

where the second equality follows from the fact that $T_S^Z$ is the first stopping time after $S$ at which $Z_\cdot^Y(m) = Z^Y$.

We finally get $T_S^Y = D_S(m)$ on the set $\{T_S^Z \leq \zeta\}$. Moreover, since $D_S(m) \leq \zeta$, this implies that $T_S^Y \leq \zeta$ on $\{T_S^Z \leq \zeta\}$ and, hence, $\{T_S^Z \leq \zeta\} = \{T_S^Y \leq \zeta\}$. Then it is immediate to see that $D_S(m) = D_{T_S^Z(m) \wedge \zeta} = T_S^Y(m) \wedge \zeta$.

Now we would like to extend the properties of $T_S^Z(m)$ w.r.t. $m$ (left-continuity and nondecreasing property) to the stopping time $T_S^Y(m)$, via the mapping $t \mapsto D_t$. It is straightforward that $m \mapsto T_S^Y(m) \wedge \zeta$ is nondecreasing since $t \mapsto D_t$ is nondecreasing. In the same way, we would immediately get the left-continuity if $t \mapsto D_t$ were left-continuous. The problem is that this last property is not true in general, but $t \mapsto D_t$ will be left-continuous along the stopping times $T_S^Z(m) \wedge \zeta$.

Let us set $D_S^-(m) := \lim \uparrow D_S(m - \epsilon)$. Thanks to the quasi-left-continuity of $Z^Y$ and $Y$ and the increasing property of $m \mapsto D_S(m)$, the preliminary remarks in Section 4.1 imply that

$$\mathbf{E}[Z_{D_S(m-\epsilon)}^Y] \to \mathbf{E}[Z_{D_S^-(m)}^Y] \quad \text{and} \quad \mathbf{E}[Y_{D_S(m-\epsilon)}] \to \mathbf{E}[Y_{D_S^-(m)}] \qquad \text{as } \epsilon \to 0.$$

Since $Z_{D_S(m-\epsilon)}^Y = Y_{D_S(m-\epsilon)}$ and $Z^Y \geq Y$, it immediately comes that $Z_{D_S^-(m)}^Y = Y_{D_S^-(m)}$ and, hence, $D_S^-(m) = D_S(m)$ since $D_S(m)$ is the first stopping time $U$ after $T_S^Z \wedge \zeta$ satisfying $Z_U^Y = Y_U$.

4. The running supremum $L_{S,u}^{Y,*}$ can be expressed as $L_{S,u}^{Y,*} = \sup\{m; T_S^Y(m) \leq u\}$. Then $m \mapsto T_S^Z(m)$ and $m \mapsto T_S^Y(m)$ define two different nondecreasing left-continuous mappings. Their right-continuous inverses $u \mapsto L_{S,u}^{Z,*}$ and $u \mapsto L_{S,u}^{Y,*}$ are therefore different but with same value at $\zeta$, thanks to the following equivalences:

$$T_S^Y(m) \leq \zeta \iff m \leq L_{S,\zeta}^{Y,*} \quad \text{and} \quad T_S^Y(m) \leq \zeta \iff T_S^Z(m) \leq \zeta.$$

Then applying Theorem 4.8, it comes that

$$C_S^{\mathrm{Am}}(Y, m) = \mathbf{E}[L_{S,\zeta}^{Z,*} \vee m | \mathcal{F}_S] - m$$
$$= \mathbf{E}[(L_{S,\zeta}^{Z,*} - m)^+ | \mathcal{F}_S] = \mathbf{E}[(L_{S,\zeta}^{Y,*} - m)^+ | \mathcal{F}_S]. \qquad \square$$

REMARK 4.12. It should be noted that the increasing processes $(L_{0,\cdot}^{Y,*})$ and $(L_{0,\cdot}^{Z,*})$ are in general different. This can be easily checked in the deterministic setting where $Z_t^Y = \sup_{t \leq u \leq T} Y_u$. The price of the American Call option can be then expressed as

$$C_t^{\mathrm{Am}}(Z^Y, m) = \sup_{t \leq u \leq T}(Z_u^Y - m)^+ = \sup_{t \leq u \leq T}\left(\sup_{u \leq v \leq T}(Y_v - m)^+\right)$$
$$= \sup_{t \leq v \leq T}(Y_v - m)^+ = C_t^{\mathrm{Am}}(Y, m) = (Z_t^Y - m)^+.$$

As for the index processes $L^Y$ and $L^Z$, they are different. In fact,

$$L_t^Z = \sup\{m; Z_t^Y(m) = Z_t^Y\} = \sup\{m; Z_t^Y \vee m = Z_t^Y\} = Z_t^Y,$$
$$L_t^Y = \sup\{m; Z_t^Y \vee m = Y_t\} = \sup\{m; Z_t^Y \vee m = Z_t^Y = Y_t\}.$$

So if $Z_t^Y = Y_t$, then $L_t^Z = L_t^Y = Z_t^Y$, otherwise $L_t^Y = -\infty$.

Let $\alpha$ define a real such that $\sup_{t \leq u \leq T} Y_u = Y_\alpha = Z_\alpha^Y = Z_0^Y$. If $\alpha \leq t$, then $L_{0,t}^{Y,*} = Z_0^Y = L_{0,t}^{Z,*}$, otherwise $L_{0,t}^{Y,*} = -\infty \neq L_{0,t}^{Z,*}$.

4.5. *Markovian case.* In a Markovian framework, we can reduce the study of all processes to that of functions.

Let $X$ be a strong Markov process ("*a right process*"), quasi-left-continuous, with lifetime $\zeta$ and topological state space $E$ whose Borel $\sigma$-field $\mathcal{B}(E)$ is separable.

The aim of the section is to reconsider the different steps of the Max-Plus decomposition in the Markovian case, in order to highlight the Markovian aspect of the different involved processes. For this, we are particularly interested in excessive functions $f$ such that $f(X_t)\mathbf{1}_{\{t < \zeta\}}$ define càdlàg supermartingales. The problem is to show that the Max-Plus decomposition of $f(X_t)\mathbf{1}_{\{t < \zeta\}}$ can be expressed through an index process $L_t = L(X_t)$. Then the problem is simply formulated as following:

Given an excessive function $f$ on $E$, does there exist any function $L$ such that

$$(4.8) \qquad f(x) = \mathbf{E}_x \left[ \sup_{0 \le t < \zeta} L(X_t) \right] = \mathbf{E}_x \left[ \oint_0^\zeta L(X_t) \right].$$

In the one-dimensional case and if $L$ is nondecreasing, equation (4.8) is equivalent to $f(x) = \mathbf{E}_x[L(\sup_{0 \le t < \zeta} X_t)]$. This kind of representation is somewhat unusual, since the classic potential theory involves a classic integral instead of a Max-Plus one. It appears, for instance, in the paper of El Karoui and Föllmer [15] who have particularly shown that any function $u$ satisfying some very mild regularity conditions admits a *nonlinear Riesz decomposition* in terms of a nonlinear potential subadditive operator $\overline{G}$ and a corresponding superadditive operator $\underline{D}$, which is a *derivator* in the sense of the nonlinear potential theory developed by Dellacherie [12]:

$$u(x) = \mathbf{E}_x \left[ \int_0^\zeta \sup_{0 \le s \le t} \underline{D} u(X_s) \, dt \right] = \overline{G} \underline{D} u(x).$$

Let $P_t f$ be the semigroup of the Markov process $X$ (with $P_t 1 \le 1$), defined by $P_t f(x) = \mathbf{E}_x[f(X_t) \mathbf{1}_{\{t < \zeta\}}]$. A function $f$ is said to be excessive if

$$f \in \mathcal{B}(E), \qquad P_t f(x) \le f(x) \quad \text{and} \quad P_t f(x) \to f(x) \qquad \text{as } t \to 0.$$

Let $\mathcal{B}^e(E)$ be the $\sigma$-field generated by the excessive functions.

With such a semigroup, the process $X$ is only sub-Markov. In order to make it a Markov process, we need to add a "cemetery" point $\partial$ to the state space $E$. We also extend all the functions defined on $E$ to null functions at the cemetery point $\partial$.

Then we introduce a realization of this Markov process on a space $(\Omega, (\mathcal{F}_t^e), \theta_t, X_t, \zeta, \mathbf{P}_x)$, with a translation operator $\theta_t$ (on $\{t < \zeta\}$, $X_t \circ \theta_s = X_{t+s}$), and a lifetime $\zeta$ ($\zeta \circ \theta_t = \zeta - t$ on $\{t < \zeta\}$, and $X_t \in \{\partial\}$ on $\{t \ge \zeta\}$). $\mathcal{F}^e$ denotes the natural filtration generated by $\mathcal{B}^e(E)$ and $X$. It is the completion of $\sigma(f(X_s); s \le t, f \in \mathcal{B}^e(E))$ with respect to the family $\{\mathbf{P}_\mu; \mu$ a finite measure on $\mathcal{P}(E)\}$.

Then for any excessive function $f$, $f(X_t) \mathbf{1}_{\{t < \zeta\}}$ is a càdlàg supermartingale for all probability measures $(\mathbf{P}_\mu; \mu \in \mathcal{P}(E))$. This allows us to avoid any continuity assumption upon $f$. Moreover, for any $g \in \mathcal{B}^e(E)$, the process $g(X_t) \mathbf{1}_{\{t < \zeta\}}$ is optional.

Let $Y_t = g(X_t) \mathbf{1}_{\{t < \zeta\}}$ be a $(\mathcal{D})$-regular process for any probability measure $(\mathbf{P}_\mu)$, where $g$ is assumed to be nonnegative. This assumption is not really required, but from the point of view of the Snell envelope and since $Y_\zeta = 0$, it is equivalent to consider $\mathrm{Sn}(Y)$ or $\mathrm{Sn}(Y \vee 0)$.

Then let us focus on the Snell envelope of $Y$. According to [14] and [18], it is associated with the function $Rg$, defined as the smallest fixed point of the operator $Kg : \mathcal{B}^e(E) \to \mathcal{B}^e(E)$, such that $Kg(x) = \sup_{r \in \mathbb{Q}} P_r g(x)$. Note that $g \geq 0 \Rightarrow Rg \geq 0$. We hence get

$$(4.9) \qquad Rg(x) = \sup_{\tau \in \mathcal{T}} \mathbf{E}_x[g(X_\tau)\mathbf{1}_{\{\tau < \zeta\}}].$$

$Rg$ can be in fact identified to the smallest excessive function that dominates $g$. If we are at time $t$, equation (4.9) becomes

$$Rg(X_t)\mathbf{1}_{\{t < \zeta\}} = \operatorname*{ess\,sup}_{\tau \in \mathcal{T}_t} \mathbf{E}_\mu[g(X_\tau)\mathbf{1}_{\{\tau < \zeta\}} | \mathcal{F}_t], \qquad \mathbf{P}_\mu\text{-a.s.,}$$

and then the addressed problem (4.8) comes down to find an optimal stopping time to the American Call option $C^{\mathrm{Am}}(Y, m)$ written on $Y$ and with maturity $\zeta$.

Note here that any constant $m$ cannot be taken as a function on $E$. In particular, to avoid any ambiguity, the corresponding constant function will be denoted by $m\mathbf{1}_E$.

THEOREM 4.13. *Let $X$ be a strong Markov process quasi-left-continuous and $v(x, m) = R(g - m)^+(x) + m\mathbf{1}_E(x)$, for $m \geq 0$. Then:*

(i) *The Max-Plus decomposition of $Rg(X_t)\mathbf{1}_{\{t < \zeta\}}$ at time $0$ is given by*

$$(4.10) \qquad Rg(x) = \mathbf{E}_x\left[\sup_{0 \leq u < \zeta} L(X_u)\right],$$

*where $L$ is the function defined on $E$ by*

$$L(x) := \sup\{m; R(g - m)^+(x) = g(x) - m\mathbf{1}_E(x)\} \qquad (\text{and } 0 \text{ elsewhere}).$$

(ii) *More generally, the price at time $0$ of the American Call $C^{\mathrm{Am}}(Y, m)$ is given by*

$$R(g - m)^+(x) = \mathbf{E}_x\left[\sup_{0 \leq u < \zeta}(L - m)^+(X_u)\right],$$

*and an optimal stopping time is characterized by*

$$T_0^Y(m) \wedge \zeta = \inf\{t \geq 0; L(X_t) \geq m\} \wedge \zeta.$$

The proof of Theorem 4.13 is simply based on the observation that the Snell envelope of $g(X_t)\mathbf{1}_{\{t < \zeta\}} \vee m$ is given by

$$v(X_t, m)\mathbf{1}_{\{t < \zeta\}} + m\mathbf{1}_{\{t \geq \zeta\}}.$$

Moreover, equation (4.10) shows that the initial problem (4.8) admits a solution, which was not clear a priori. However this solution is not necessarily unique, since there is no uniqueness for the nondecreasing processes involved in the Max-Plus decomposition of $Rg(X_t)\mathbf{1}_{\{t < \zeta\}}$.

A direct discussion of the Markovian case and further results concerning the uniqueness of $L$ can be found in Föllmer and Knispel [26].

REMARK 4.14. At time $t$, the Max-Plus decomposition of $Rg(X_t)\mathbf{1}_{\{t<\zeta\}}$ becomes

$$Rg(X_t)\mathbf{1}_{\{t<\zeta\}} = \mathbf{E}_x\left[\sup_{t\le u<\zeta} L(X_u)|\mathcal{F}_t\right],$$

where $x$ denotes the initial value of the underlying process $(X_t)$, and the American Call price at time $t$ is given by

$$R(g-m)^+(X_t) = \mathbf{E}_x\left[\sup_{t\le u<\zeta}(L-m)^+(X_u)|\mathcal{F}_t\right].$$

EXAMPLE. Let us come back to the example of Section 3 and reinterpret it in terms of a Markov process. The supermartingale $Z$ defines a one-dimensional Markov process evolving according to a geometric Brownian motion

$$\frac{dZ_t}{Z_t} = -r\,dt + \sigma\,dW_t, \qquad Z_0 = 1.$$

Let us introduce a lifetime $\zeta$ assumed to be an independent exponential variable with parameter $\beta > 0$ and a cemetery point $\partial$ (we set $Z_t = \partial$, if $t \ge \zeta$). $Z$ is hence extended to a Markov process with values in the enlarged space $\mathbb{R} \cup \{\partial\}$.

Then we formally define the identity function by $Id(x) = x$ on $\mathbb{R}$ and $Id(\partial) = 0$. Note that, with this convention, the supermartingale $\widetilde{Z}_t = Z_t\mathbf{1}_{\{t<\zeta\}}$ introduced in Section 3 can be written as $\widetilde{Z}_t = Id(Z_t)$.

Thanks to Theorem 4.13, the Snell envelope $\widetilde{Z}.(m)$ of $\widetilde{Z} \vee m$ is a function of $(Z,m)$ and can be decomposed as following in the Max-Plus algebra:

$$\widetilde{Z}_t(m) = \mathbf{E}\left[\sup_{t\le u<\zeta} L(Z_u)\mathbf{1}_{\{u<\zeta\}} \vee m|\mathcal{F}_t\right] \qquad \text{for } t<\zeta, \text{ and } m \text{ for } t\ge\zeta,$$

where $L(x) = \frac{\delta-1}{\delta}x$, as we have seen in Section 3 (cf. Lemma 3.5 and Example 3.6). We finally obtain

$$\widetilde{Z}_t = \mathbf{E}\left[\sup_{t\le u\le\zeta}\frac{\delta-1}{\delta}Z_u\mathbf{1}_{\{u<\zeta\}}|\mathcal{G}_t\right] = \frac{\delta-1}{\delta}\mathbf{E}[\widetilde{Z}^*_{t,\zeta}|\mathcal{G}_t].$$

## 5. Optimality of Max-Plus decomposition w.r.t. convex order.

For the sake of completeness, we first give some useful definitions and properties of stochastic order, which expresses the notion of one entire probability distribution being less than or equal to another. This order which was introduced in Economics by Rothschild and Stiglitz [43] as a measure of risk gives a systematic framework for analyzing economic behavior under uncertainty.

More generally, stochastic order relations provide a valuable insight into the behavior of complex stochastic systems. Application areas include queuing systems, actuarial and financial risk, decision making and stochastic simulation.

5.1. *Basic properties of convex order.*

DEFINITION 5.1.  Let $X_1$ and $X_2$ be two real-valued random variables. We say that $X_1$ is less variable than $X_2$ in the *convex stochastic order*, and we write $X_1 \leq_{cx} X_2$ if for any convex real-valued function $g$ for which the following expectations are well defined:

$$(5.1) \qquad \qquad \mathbf{E}[g(X_1)] \leq \mathbf{E}[g(X_2)].$$

When $\mathbf{E}[X_1] = \mathbf{E}[X_2]$, the test functions can be reduced to $\Phi_m(x) = x \vee m$ for all reals $m$.

Let us point out some basic facts concerning stochastic order:

• The convex order compares the "dispersion" of random variables with equal mean. In particular, by considering specific convex functions, it can be easily seen that

$$X_1 \leq_{cx} X_2 \Rightarrow \mathrm{Var}(X_1) \leq \mathrm{Var}(X_2) \qquad \text{whenever } \mathrm{Var}(X_2) < \infty.$$

However, the converse implication does not hold. So, indeed, the convex ordering is stronger than ordering of the variances since it takes into account irregular or asymmetric risky prospects.

• Recall that a real-valued function $g$ is *convex* if $g(\mathbf{E}[X]) \leq \mathbf{E}[g(X)]$ for all r.v. $X$. We particularly get that $\mathbf{E}[X] \leq_{cx} X$ for all r.v. $X$. More generally, note that if $g$ is a convex function, then, by Jensen's inequality,

$$\mathbf{E}\{g(\mathbf{E}[X_2|X_1])\} \leq \mathbf{E}\{\mathbf{E}[g(X_2)|X_1]\} = \mathbf{E}[g(X_2)].$$

This means that $\mathbf{E}[X_2|X_1] \leq_{cx} X_2$. This property is in fact characteristic in the sense of the next beautiful *Strassen's theorem*, which characterizes the convex order by construction on the same probability space:

*If $X_1$ and $X_2$ are two r.v. such that $X_1 \leq_{cx} X_2$, then there exist two r.v. $\widetilde{X_1}$ and $\widetilde{X_2}$ defined on a common probability space such that*

$$\widetilde{X_i} \stackrel{d}{=} X_i \qquad \text{for } i = 1, 2 \quad \text{and} \quad \widetilde{X_1} = \mathbf{E}[\widetilde{X_2}|\widetilde{X_1}] \qquad a.s.$$

This theorem on stochastic dominance is a crucial tool in the theory of interacting particle systems, and has also found many interesting applications in other areas. See [27] (Corollary 2.100) or [40] for a proof of the theorem.

• Note that $X_1 \leq_{cx} X_2$ is equivalent to $-X_1 \leq_{cx} -X_2$. This means that, contrary to the monotone convex orders, the convex order is independent of the interpretation of the random variables as loss or gain variables.

• It is said that $X_1$ is smaller than $X_2$ in the *decreasing convex order*, written $X_1 \leq_{dcx} X_2$, if inequality (5.1) holds for all decreasing convex functions $g$, for which the expectations exist.

This particularly implies that $\mathbf{E}[X_1] \geq \mathbf{E}[X_2]$ and if the two means are equal, then the decreasing convex order reduce to the convex order.

• Note that the decreasing convex order is strictly equivalent to the classic second order stochastic dominance relation in finance, which is the increasing concave order. It is a fundamental model of risk-averse preferences and has an equivalent characterization by utility functions. The term "decreasing convex" is natural when we are dealing with minimization rather than maximization problems. In fact, von Neumann and Morgenstern have developed the expected utility theory [39]: for every rational decision maker, there exists a utility function $u(\cdot)$ such that the decision maker prefers outcome $X$ over outcome $Y$ if and only if $\mathbf{E}[u(X)] > \mathbf{E}[u(Y)]$. In practice, however, it is almost impossible to explicitly elicit the utility function of a decision maker. Additional difficulties arise when there is a group of decision makers with different utility functions who have to come to a consensus.

In terms of utility theory, $X \leq_{dcx} Y$ means that $\mathbf{E}[u(X)] \geq \mathbf{E}[u(Y)]$ for every nondecreasing and concave utility function $u(\cdot)$, that is, the gain $X$ is preferred to the gain $Y$ by all risk averse decision makers.

We refer the reader to the book by Shaked and Shanthikumar [44] for an overview of the convex order and other stochastic orders.

5.2. *Martingale optimization problems.* It is natural to address the problem to "find the smallest martingale" dominating a floor process in finance, where martingales may be seen as self-financing portfolio strategies, and also in fair games language, where martingales may be thought of as the fortune earned by a betting strategy.

However, the set of martingales is not stable with respect to the infimum operation, since the inf of two martingales defines a supermartingale and not a martingale. Hence, the problem has no solution in general and we have to weaken the assumption of the "strong order" to a convex stochastic order. While this kind of problem, set in terms of convex order, is somewhat unusual in finance since we consider in general only one convex function, it seems to be more classic in other areas of probability theory. For instance, Kertz and Rösler have addressed in [33] a martingale problem similar to ours, in which the optimization is also related to the convex stochastic order on terminal values. However, our domination path constraint is replaced by a constraint imposing that the distribution of the maximum of the martingale is a given probability measure $\nu$. The solution is thoroughly characterized, using the notions of Hardy–Littlewood maximal functions and convex envelopes.

Let us formulate our "new" constrained optimization problem in terms of convex order. We use the same notation as in Section 4.4: $Z^Y$ is the Snell envelope of a real-valued optional process $Y$ of class $(\mathcal{D})$.

Introduce the following set of admissible martingales:

$$\mathcal{M}^Y = \{(M_t)_{t \geq 0} \text{ u.i. martingale} \mid M_0 = Z_0^Y \text{ and } M_t \geq Y_t \ \forall t \in [0, \zeta]\}.$$

Note that any martingale dominating a floor process $Y$ necessarily dominates its Snell envelope $Z^Y$ and, thus, in order to satisfy the floor constraint $M_t \geq Y_t$ for all $t \leq \zeta$, the initial value of any admissible martingale $M$ must be at least equal to the initial value of the Snell envelope of $Y$, that is, $M_0 \geq Z_0^Y = \sup_{\tau \in \mathcal{T}_{0,\zeta}} \mathbf{E}[Y_\tau]$.

*Our aim is to find the smallest martingale $M^*$ in $\mathcal{M}^Y$ with respect to the convex stochastic order on the terminal value, that is, $M_\zeta^* \leq_{cx} M_\zeta$ for all martingales $(M_t)_{0 \leq t \leq \zeta}$ in $\mathcal{M}^Y$.*

First, it is easy to check that the set of admissible martingales $\mathcal{M}^Y$ is not empty since it already contains the martingale $M^A(Y)$ of the Doob–Meyer decomposition of $Z^Y$.

Moreover, it is shown in [42] that this martingale $M^A(Y)$ achieves the minimum over all martingales $M$ with initial value $Z_0^Y$, in the following representation:

$$(5.2) \qquad Z_0^Y = \inf_M \mathbf{E}\left[\sup_{t \in [0,\zeta]} (Y_t - M_t)\right] + M_0.$$

However, it is also trivially the case for all admissible martingales in $\mathcal{M}^Y$ since

$$\mathbf{E}\left[\sup_{t \in [0,\zeta]} (Y_t - M_t)\right] + M_0 \leq \mathbf{E}\left[\sup_{t \in [0,\zeta]} (Z_t^Y - M_t)\right] + M_0 \leq M_0 = Z_0^Y$$
$$\forall M \in \mathcal{M}^Y.$$

The addressed problem is in general difficult to solve and usually we just consider the Doob–Meyer martingale $M^A(Y)$.

The following theorem states that the martingale $M^{Y,\oplus}$, introduced in Section 4.4, solves our constrained optimization problem, and then, in particular, $M_\zeta^{Y,\oplus}$ is less variable than $M_\zeta^A(Y)$. We still assume in the theorem that the filtration $(\mathcal{F}_t)$ is quasi-left- continuous.

THEOREM 5.2. *The martingale $M^{Y,\oplus}$ of the Max-Plus decomposition of $Z^Y$ is the smallest martingale in $\mathcal{M}^Y$, with respect to the convex stochastic order on the terminal value.*

PROOF. Let $(M_t)_{0 \leq t \leq \zeta}$ be an arbitrary element of $\mathcal{M}^Y$ and $(L_{0,t}^*)$ the nondecreasing process in the Max-Plus decomposition of $Z^Y$. We shall prove that $M_\zeta^{Y,\oplus} \leq_{cx} M_\zeta$.

Since $M$ dominates $Z^Y$, the Snell envelope $Z^M(m)$ of $(M \vee m)$ also dominates $(Z.(m))$.

However, since we have previously observed that $Z_S^M(m) = \mathbf{E}[M_\zeta \vee m | \mathcal{F}_S]$ for all $S$ in $\mathcal{T}$, it immediately follows that

$$\mathbf{E}[M_\zeta \vee m | \mathcal{F}_S] \geq \mathbf{E}[L_{S,\zeta}^* \vee Z_\zeta^Y \vee m | \mathcal{F}_S] \qquad \forall S \in \mathcal{T}.$$

More generally, this inequality holds true for any convex function $g$ that is

$$(5.3) \qquad \mathbf{E}[g(M_\zeta)|\mathcal{F}_S] \geq \mathbf{E}[g(L^*_{S,\zeta} \vee Z^Y_\zeta)|\mathcal{F}_S] \qquad \forall S \in \mathcal{T}.$$

The terminal condition is $M^\oplus_\zeta = L^*_{0,\zeta} \vee Z^Y_\zeta$, and so equation (5.3) implies $\mathbf{E}[g(M_\zeta)] \geq \mathbf{E}[g(L^*_{0,\zeta} \vee Z^Y_\zeta)] \geq \mathbf{E}[g(M^{Y,\oplus}_\zeta)]$ and the martingale $(M^{Y,\oplus}_t)_{0 \leq t \leq \zeta}$ is optimal indeed.

It should be noted that the same argument can only be used for stopping times $S$ such that $L^*_{0,S} = L_S$, that is at the increasing points of $L^*_{0,\cdot}$. □

Observe that:

- If we impose an other initial value $m$ for all admissible martingales in $\mathcal{M}^Y$, $m$ must be necessarily greater than $Z^Y_0$, otherwise the problem has no solution. Then the same results hold if we replace the increasing process $(L^{Y,*}_{0,t})_{t \geq 0}$ by $(L^{Y,*}_{0,t} \vee m)_{t \geq 0}$.

- Since the initial value of any martingale is equal to its mean, the formulation of the initial condition strongly depends on the selected stochastic order. If we consider the convex order, all admissible martingales must have the same initial value $m$, whereas if we consider the decreasing convex order, the initial value of any admissible martingale $M$ must not exceed the initial value of any optimal solution to the problem. It must also be equal or greater than $Z^Y_0$. Note that while the original optimization problem was difficult to solve, the dual one (passing through the Snell envelope of a convex family of processes and its decomposition in the Max-Plus algebra) is much simpler than the first one and requires only tools from convex analysis.

REMARK 5.3. If we consider a new financial environment in which portfolio strategies are martingales, the problem treated here can be applied to portfolio insurance. In fact, it can be seen as a particular portfolio selection problem where an American constraint is imposed on the liquidative value of the open fund. Traditionally, investors are assumed to possess an increasing concave utility function $u$ and the portfolio choice consists in maximizing the expected utility of final wealth, over the set of admissible portfolios. In practice, however, it is almost impossible to explicitly elicit the utility function of a decision maker.

In our approach, the optimization is performed with respect to the convex ordering on the terminal value and the optimal strategy is thus robust to different preferences. This model is hence very useful, especially when there is a group of decision makers with different utility functions who have to come to a consensus. The interested reader is referred to [20] for more details on the application of the martingale problem to portfolio insurance.

5.3. *Max-Plus martingales and Lévy processes.* As we have seen in Section 3, processes with independent stationary increments give us nice examples where the Max-Plus decomposition can be described in quasi-closed form. While the focus of Section 3 was the link to American options and characterization of optimal stopping times, we rather emphasize here the different martingales involved in the Max-Plus decomposition of Lévy processes. The interesting distinction to make here is between infinite horizon time and an independent exponential random variable one. We consider both cases and determine in each one the closed form and the dynamics of the Max-Plus martingale $M^\oplus$.

5.3.1. *Analytical result.* Let $Z$ be a geometric Brownian motion with parameters $(-r, \sigma)$ and initial value $Z_0 = x$. We just recall some classic results on the probability distribution of the running supremum of $Z$, that is, $Z_t^* := Z_{0,t}^* = \sup_{0 \le u \le t} Z_u$.

LEMMA 5.4. *Set $\gamma = 1 + \frac{2r}{\sigma^2}$ and $\delta$ the root greater than $\gamma$ of $y^2 - \gamma y - \frac{2\beta}{\sigma^2} = 0$. Let $\zeta$ be an independent exponential variable with parameter $\beta > 0$. The case where $\zeta$ is infinite a.s. corresponds to $\beta = 0$ and will be considered together.*

1. $\mathbf{P}[Z_\zeta^* \ge m] = \left(\dfrac{x}{m} \wedge 1\right)^\delta$ *and* $\mathbf{E}[Z_\zeta^*] = \dfrac{\delta}{\delta - 1} x$.

2. $\mathbf{E}[(Z_\zeta^* - m)^+] = \begin{cases} \dfrac{m}{\delta - 1} \left(\dfrac{x}{m}\right)^\delta, & \text{if } m \ge x \\ \dfrac{\delta}{\delta - 1} x - m, & \text{if } m \le x \end{cases} = C_\delta^*(x, m).$

PROOF. 1. The proof is based on the equivalence $\{Z_t^* \ge m \Leftrightarrow T_m \le t\}$, where $T_m := \inf\{t; Z_t^* \ge m\}$. So, $\mathbf{P}[Z_\zeta^* \ge m] = \mathbf{E}[\exp -\beta T_m]$ and even if $\beta = 0$, $\mathbf{P}[Z_{0,\infty}^* \ge m] = \mathbf{P}[T_m < +\infty] = \lim_{\beta \to 0} \mathbf{E}[\exp -\beta T_m]$.

To calculate these quantities, we apply Doob's theorem to the well-selected martingale, $e^{-\beta t} Z_t^\delta$, where $\delta$ is the positive root of the equation $\delta^2 - \gamma \delta - \frac{2\beta}{\sigma^2} = 0$, $(\delta = \gamma$ if $\beta = 0)$. So, $\mathbf{E}[e^{-\beta(T_m \wedge t)} Z_{T_m \wedge t}^\delta] = x^\delta$. Since $e^{-\beta(T_m \wedge t)} Z_{T_m \wedge t}^\delta$ is nonnegative and bounded by $m \vee x$ for all $t$, when $t \to +\infty$, the left-hand side goes to $(m \vee x)^\delta \times \mathbf{E}[e^{-\beta T_m}] = (m \vee x)^\delta \times \mathbf{P}[Z_\zeta^* \ge m]$. This proves item 1.

2. More generally, for $m \ge x$, the price of a Call option on the supremum is given by

$$\mathbf{E}[(Z_\zeta^* - m)^+] = \int_0^{+\infty} \mathbf{P}[Z_\zeta^* - m \ge \alpha] \, d\alpha = \int_0^{+\infty} \left(\frac{x}{\alpha + m}\right)^\delta d\alpha = \frac{m}{\delta - 1} \left(\frac{x}{m}\right)^\delta.$$

But if $m \le x$, $Z_\zeta^* \ge m$ and $\mathbf{E}[(Z_\zeta^* - m)^+] = \mathbf{E}[Z_\zeta^* - m] = \frac{\delta}{\delta - 1} x - m$. □

This lemma proves Proposition 3.4. We first assume the maturity $\zeta$ to be infinite.

5.3.2. *Infinite horizon. Multiplicative Lévy processes.* Let $Z$ be a multiplicative Lévy process such that $\mathbf{E}[Z^*_{0,\infty}] < +\infty$. The Max-Plus decomposition of $Z$ has a closed form thanks to the independence property of its relative increments.

*Geometric Brownian motion.*

PROPOSITION 5.5. *Let $Z$ be a geometric Brownian motion with parameters $(-r, \sigma)$, and set $\gamma = 1 + \frac{2r}{\sigma^2}$. The martingale of the Max-Plus decomposition of $Z$ can be explicitly characterized as a function $\phi_\gamma$ of $(Z_t, Z^*_t)$:*

$$M^\oplus_t = \frac{\gamma - 1}{\gamma} Z^*_t \left[ \frac{1}{\gamma - 1} \left( \frac{Z_t}{Z^*_t} \right)^\gamma + 1 \right] := \phi_\gamma(Z_t, Z^*_t).$$

*In particular,*    $M^\oplus_t = Z_t \oplus L^*_{0,t}$    *if and only if*    $Z_t = Z^*_t$.

*Moreover, as a martingale, $M^\oplus_t$ can be represented as a stochastic integral*

$$dM^\oplus_t = \left( \frac{Z_t}{Z^*_t} \right)^{\gamma - 1} \sigma Z_t \, dW_t.$$

PROOF.   By the uniqueness Theorem 2.8, $M^\oplus_t = b_\gamma \mathbf{E}[Z^*_{0,\infty} | \mathcal{F}_t]$ is the martingale of the Max-Plus decomposition of $Z$. Since the distribution of $Z^*_{0,\infty}$ is well known, there exists a closed formula for $M^\oplus_t$ as a function of $(Z_t, Z^*_t)$:

$$(5.4) \qquad \begin{aligned} M^\oplus_t &= b_\gamma \mathbf{E}[(Z^*_{t,\infty} - Z^*_t)^+ | \mathcal{F}_t] + b_\gamma Z^*_t \\ &= b_\gamma Z_t \mathbf{E}\left[ \left( \frac{Z^*_{t,\infty}}{Z_t} - \frac{Z^*_t}{Z_t} \right)^+ \Big| \mathcal{F}_t \right] + b_\gamma Z^*_t. \end{aligned}$$

Since $Z^*_{t,\infty}/Z_t$ is independent from $Z^*_t/Z_t$ and has the same distribution as $Z^*_{0,\infty}/x$, equation (5.4) can be rewritten as follows:

$$(5.5) \qquad M^\oplus_t = b_\gamma Z_t C^*_\gamma(1, m_t) + b_\gamma Z^*_t, \qquad \text{with } m_t := \frac{Z^*_t}{Z_t} \geq 1.$$

Then substituting $m_t$ in the last expression of Lemma 5.4, we explicitly determine $M^\oplus$:

$$M^\oplus_t = \frac{\gamma - 1}{\gamma} Z^*_t \left[ \frac{1}{\gamma - 1} \left( \frac{Z_t}{Z^*_t} \right)^\gamma + 1 \right] := \phi_\gamma(Z_t, Z^*_t).$$

Since the associated nondecreasing process is in the form of $L^*_{0,t} = b_\gamma Z^*_t = \frac{\gamma - 1}{\gamma} Z^*_t$, it can be immediately seen that the martingale $M^\oplus$ is different from $Z \vee L^*_{0,\cdot}$:

$$M^\oplus_t - (Z_t \vee L^*_{0,t}) = \frac{\gamma - 1}{\gamma} Z^*_t \left[ \frac{1}{\gamma - 1} \left( \frac{Z_t}{Z^*_t} \right)^\gamma - \left( \frac{\gamma}{\gamma - 1} \frac{Z_t}{Z^*_t} - 1 \right)^+ \right].$$

The function $x \mapsto \frac{1}{\gamma-1}x^\gamma - (\frac{\gamma}{\gamma-1}x - 1)^+$ is nonnegative on $[0,1]$, null at 0 and 1 and reaches its maximum at the point $b_\gamma = \frac{\gamma-1}{\gamma}$.

Then $M_t^\oplus = Z_t \vee L_{0,t}^*$ if and only if $Z_t = Z_t^*$, that is, if and only if $t$ is a point of increase of the process $L_{0,\cdot}^*$.

In addition, since $M^\oplus = \phi_\gamma(Z, Z^*)$ is a martingale, by Itô's formula, its decomposition as a stochastic integral needs only to know the derivative of $\phi_\gamma$ w.r.t. $Z$: $dM_t^\oplus = \frac{\partial \phi_\gamma}{\partial x}(Z_t, Z_t^*)Z_t \sigma \, dW_t$. We hence obtain

$$(5.6) \qquad dM_t^\oplus = \left(\frac{Z_t}{Z_t^*}\right)^{\gamma-1} \sigma Z_t \, dW_t = \left(\frac{Z_t}{Z_t^*}\right)^{\gamma-1} dM_t^A,$$

where $M^A$ denotes the martingale of the Doob–Meyer decomposition of $Z$. From (5.6), we can particularly observe that $M^\oplus$ is less variable than $M^A$. □

*Geometric Lévy process.* Let $C^{\mathrm{Am}}(Z, m)$ be a perpetual American Call written on a geometric Lévy process $Z$ defining a supermartingale, and with strike $m$. The main difficulty to compute its price comes from the complexity of Lookback options closed formulae. Based on the paper [38], we have explicitly determined in Section 3.2 the price of such a Call, in the particular case where $Z_t = xe^{X_t}$, with $X$ defining an upper semi-continuous Lévy process (i.e., a process with no positive jumps). Then using the relation $M_t^\oplus = C_t^{\mathrm{Am}}(Z, bZ_t^*) + bZ_t^*$ which follows from equation (5.4) and Proposition 3.2, we easily deduce a closed formula for the Max-Plus martingale $M^\oplus$ as a function $\phi_{\gamma_{\mathrm{Lévy}}}$ of $(Z_t, Z_t^*)$. $\phi_{\gamma_{\mathrm{Lévy}}}$ has the same form as in Proposition 5.5 and $\gamma_{\mathrm{Lévy}} > 1$ is such that $\kappa(\gamma_{\mathrm{Lévy}}) = 0$.

*Brownian motion with negative drift.* Let $Z$ define a drifted Brownian motion with parameters $(-\mu, \sigma)$, where $\mu = r + \frac{\sigma^2}{2}$ and initial value 0, and assume the maturity $\zeta$ to be infinite. The martingale of the Max-Plus decomposition of $Z$ has a closed formula based on the independence property of the increments of $Z$. We have seen in Section 3.3 that $Z_t$ takes the simple following form: $Z_t = \mathbf{E}[Z_{t,\infty}^* - b|\mathcal{F}_t]$, where $b$ is set to be equal to $\mathbf{E}[Z_{0,\infty}^*]$. Then by the uniqueness theorem 2.8, $M_t^\oplus = \mathbf{E}[Z_{0,\infty}^* - b|\mathcal{F}_t]$ is the martingale of the Max-Plus decomposition associated with the running supremum of the process $Z_t - b$. The law of $Z_{0,\infty}^*$ can be deduced from Lemma 5.4, and so we obtain a closed formula for $M_t^\oplus$ as a function of $(Z_t, Z_{0,t}^*)$.

The following proposition provides an explicit characterization of $M^\oplus$. The proof is omitted here since it is strictly analogous to that of Proposition 5.5.

PROPOSITION 5.6. *The martingale $M^\oplus$ associated with the Max-Plus decomposition of $Z$ is of the form*

$$(5.7) \qquad M_t^\oplus = \frac{1}{\gamma}[\exp(-\gamma(Z_t^* - Z_t)) - 1] + Z_t^* := \phi(Z_t, Z_t^*).$$

In particular, $M_t^\oplus = Z_t \oplus L_{0,t}^*$ if and only if $Z_t = Z_t^*$.

Moreover, as a martingale, $M_t^\oplus$ can be represented as a stochastic integral

$$dM_t^\oplus = \exp(-\gamma(Z_t^* - Z_t))\sigma \, dW_t = \exp(-\gamma(Z_t^* - Z_t)) \, dM_t^A.$$

5.3.3. *Independent exponential r.v. horizon.* Now, we assume that the maturity $\zeta$ defines an independent exponential r.v. with parameter $\beta > 0$, and $Z$ a geometric Brownian motion with parameters $(-r, \sigma)$. We use the same notation as in Section 3.2.2.

The martingale $\widetilde{M}^\oplus$ associated with the Max-Plus decomposition of $\widetilde{Z}$ is of the form

$$\widetilde{M}_t^\oplus = b_\beta \mathbf{E}\left[\sup_{0 \leq u \leq \zeta} \widetilde{Z}_u | \mathcal{G}_t\right] = b_\beta \mathbf{E}[\widetilde{Z}_{0,\zeta}^* | \mathcal{G}_t] = b_\beta \mathbf{E}[Z_{0,\zeta}^* | \mathcal{G}_t].$$

Since the distribution of $Z_{0,\zeta}^*$ is also explicit in this case, we can again deduce a closed formula of $\widetilde{M}^\oplus$ as a function of $(Z_t, Z_{0,t}^*)$, for $t < \zeta$.

Note that the law of $Z_{0,\zeta}^*$ has the same form as within the infinite time horizon, but $\gamma$ is replaced by $\delta$. Then, the calculations are strictly identical and lead to the same closed form for the martingale $\widetilde{M}^\oplus$.

PROPOSITION 5.7. 1. *The Max-Plus martingale associated with* $\widetilde{Z}_t = Z_t \mathbf{1}_{\{t < \zeta\}}$ *is of the form*

$$\widetilde{M}_t^\oplus = \frac{\delta - 1}{\delta} \widetilde{Z}_t^* \left[\frac{1}{\delta - 1}\left(\frac{\widetilde{Z}_t}{\widetilde{Z}_t^*}\right)^\delta + 1\right] = \phi_\delta(\widetilde{Z}_t, \widetilde{Z}_t^*) \qquad \text{for all } t \in [0, \zeta].$$

*In particular,* $\widetilde{M}_t^\oplus = \widetilde{Z}_t \oplus \widetilde{L}_{0,t}^*$ *if and only if* $\widetilde{Z}_t = \widetilde{Z}_t^*$.

2. *As a martingale,* $\widetilde{M}_t^\oplus$ *can be represented as the sum of a stochastic integral and a purely discontinuous martingale, for all* $t \leq \zeta$,

$$d\widetilde{M}_t^\oplus = \left(\frac{Z_t}{Z_t^*}\right)^{\delta - 1} \mathbf{1}_{\{t < \zeta\}} \, dM_t^A + \Delta \widetilde{M}_{t^-}^\oplus \, d\widetilde{N}_t^d,$$

*where* $\widetilde{N}_t^d := \mathbf{1}_{\{\zeta \leq t\}} - \beta(t \wedge \zeta)$ *and* $\Delta \widetilde{M}_{\zeta^-}^\oplus = \widetilde{M}_\zeta^\oplus - \widetilde{M}_{\zeta^-}^\oplus = -\frac{Z_\zeta^*}{\delta}\left(\frac{Z_\zeta}{Z_\zeta^*}\right)^\delta$.

PROOF. 1. For any $t < \zeta$, $\widetilde{M}_t^\oplus = \frac{\delta - 1}{\delta} Z_t \mathbf{E}[m_t \vee \frac{Z_{t,\zeta}^*}{Z_t} | \mathcal{G}_t]$, where $m_t = \frac{Z_{0,t}^*}{Z_t}$. Since $\frac{Z_{t,\zeta}^*}{Z_t}$ is conditionally independent given $\mathcal{G}_t$ and distributed like $\frac{Z_{0,\zeta}^*}{x}$, then, applying Lemma 5.4, we obtain $\widetilde{M}_\zeta^\oplus = \frac{\delta - 1}{\delta} Z_\zeta^*$, and for $t < \zeta$,

$$\begin{aligned}
\widetilde{M}_t^\oplus &= \frac{\delta - 1}{\delta} Z_t \mathbf{E}\left[m_t \vee \frac{Z_{0,\zeta}^*}{x}\right] = \frac{\delta - 1}{\delta} \frac{Z_t}{x}(\mathbf{E}[(Z_\zeta^* - m_t x)^+] + m_t x) \\
&= \frac{\delta - 1}{\delta} \frac{Z_t}{x}\left(\frac{m_t x}{\delta - 1}\left(\frac{1}{m_t}\right)^\delta + m_t x\right) = \frac{\delta - 1}{\delta} Z_t^*\left[\frac{1}{\delta - 1}\left(\frac{Z_t}{Z_t^*}\right)^\delta + 1\right].
\end{aligned}$$

2. At time $\zeta$, the process $\widetilde{Z}$ jumps from $\widetilde{Z}_{\zeta^-} = Z_\zeta$ to 0. So, the martingale $\widetilde{M}_\zeta^\oplus$ has also a jump of size

$$\Delta\widetilde{M}_\zeta^\oplus = \widetilde{M}_\zeta^\oplus - \widetilde{M}_{\zeta^-}^\oplus = \phi_\delta(\widetilde{Z}_\zeta, \widetilde{Z}_\zeta^*) - \phi_\delta(Z_\zeta, Z_\zeta^*)$$
$$= -\frac{Z_\zeta^*}{\delta}\left(\frac{Z_\zeta}{Z_\zeta^*}\right)^\delta = \frac{1}{\delta}\left(\frac{Z_\zeta}{Z_\zeta^*}\right)^{\delta-1}\Delta\widetilde{Z}_\zeta.$$

Since the process $\widetilde{N}_t^d := \mathbf{1}_{\{\zeta \le t\}} - \int_0^{t\wedge\zeta}\beta\,du$ defines a martingale and $\Delta\widetilde{M}_\zeta^\oplus$ is the value at $\zeta$ of a $(\mathcal{G}_t)$-predictable process, then

$$\widetilde{M}_t^{\oplus,d} := \Delta\widetilde{M}_\zeta^\oplus\mathbf{1}_{\{\zeta \le t\}} + \int_0^{t\wedge\zeta}\beta\frac{Z_u^*}{\delta}\left(\frac{Z_u}{Z_u^*}\right)^\delta du$$

is the purely discontinuous part of the $\mathcal{G}$-martingale $\widetilde{M}_t^\oplus$.

Observe that if $\widetilde{N}_t^{\widetilde{Z},d} := -Z_\zeta\mathbf{1}_{\{\zeta \le t\}} + \int_0^{t\wedge\zeta}\beta Z_u\,du$ is the martingale corresponding to the jump of $\widetilde{Z}$, then

$$d\widetilde{M}_t^{\oplus,d} = \frac{1}{\delta}\left(\frac{Z_t}{Z_t^*}\right)^{\delta-1}d\widetilde{N}_t^{\widetilde{Z},d}.$$

The continuous part of the martingale $\widetilde{M}_t^\oplus$ is given by $\widetilde{M}_t^{\oplus,c} = \phi_\delta(Z_{t\wedge\zeta}, Z_{t\wedge\zeta}^*) - \int_0^{t\wedge\zeta}\beta\frac{Z_u^*}{\delta}(\frac{Z_u}{Z_u^*})^\delta du$ with infinitesimal decomposition

$$d\widetilde{M}_t^{\oplus,c} = \frac{\partial\phi_\delta}{\partial x}(Z_{t\wedge\zeta}, Z_{t\wedge\zeta}^*)\,dM_{t\wedge\zeta}^A,$$

where $M^A$ denotes the martingale of the Doob–Meyer decomposition of $Z$. $\square$

Note that in the infinite time horizon case, $M^\oplus$ was continuous even in $\zeta = \infty$, since $Z_t$ vanishes as $t \to \infty$. But here $Z_\zeta$ does not a priori equal $\widetilde{Z}_\zeta = 0$, and thus, the martingale $\widetilde{M}^\oplus$ jumps at time $\zeta$.

### 5.4. *Azéma–Yor martingale and Max-Plus decomposition.*

We leave in this section the framework of Lévy processes and focus on a new example of Max-Plus decomposition where we can make explicit calculations with no assumption of stationary independent increments.

So let $Z$ be an increasing concave function $u$ of a continuous local martingale $N$ that goes to 0 as $t \to +\infty$. Let us also assume the time horizon $\zeta$ to be infinite. Thanks to the concavity property of $u$, $Z_t = u(N_t)$ defines a local supermartingale. We further assume that $\mathbf{E}[|u(\sup_{t\ge 0} N_t)|] < +\infty$ so that $Z$ is of class $(\mathcal{D})$. Then setting $N_t^* = \sup_{0 \le s \le t} N_s$ and applying Itô's formula in the case where $u$ is in the class $\mathcal{C}^1$, we observe that the process

$$(5.8) \quad M_t := u(N_t^*) + (N_t - N_t^*)u'(N_t^*) = u(N_0) + \int_0^t u'(N_s^*)\, dN_s$$

defines a local martingale. This key property has been greatly used by Azéma–Yor without any concavity assumption, and particularly served to solve the Skorohod problem in [5]. More generally, the same property still holds if $N_t$ jumps but not $N_t^*$, which is the case, for example, for any positive martingale with only negative jumps.

5.4.1. *Max-Plus martingale and Azéma–Yor martingale.* Since $u$ is concave, it comes that $u(y) - u(x) \le u'(x)(y - x)$ for all reals $(x, y)$, and so the martingale $M_t$ dominates $Z_t = u(N_t)$.

We aim to apply the uniqueness Theorem 2.8 in order to prove that $M$ is the martingale associated with the Max-Plus decomposition of $Z$. For this, $M_\infty$ must equal the terminal value of a nondecreasing process $L_{0,\cdot}^*$, which only increases at points in time $t$ such that $L_{0,t}^* = L_t$.

As we have assumed that $N_\infty = 0$, it immediately follows that $N_\infty^* \ge 0$ and

$$M_\infty = \lim_{t \to +\infty} M_t = u(N_\infty^*) - u'(N_\infty^*)N_\infty^* := v(N_\infty^*).$$

Since $v'(x) = -xu''(x)$ if $u$ is regular and thanks to the concavity of $u$, $v'(x) \ge 0$ if $x \ge 0$ and the function $v$ is nondecreasing on $[0, +\infty)$. Moreover, $N_\infty^* = (N^+)_\infty^*$ since $N_\infty^* \ge 0$, but we have not necessarily $N_t^* = (N^+)_t^*$ for any time $t$. Hence, $M_\infty$ is the terminal value of a nondecreasing process $v((N^+)_t^*) = \sup_{0 \le s \le t} L_s$, where $L_s := u(N_s^+) - u'(N_s^+)N_s^+$. In addition, while $N_t \le 0$, $(N^+)_t^*$ remains sticked to 0 and does not increase. So the nondecreasing process $v((N^+)_t^*)$ only increases when $(N^+)_t^* = N_t^+ = N_t = N_t^*$, that is, $M_t = Z_t$ [in light of equation (5.8)].

Consequently, $M$ satisfies all the conditions of Theorem 2.8, which ensures the uniqueness of the martingale $M = M^\oplus$ associated with the Max-Plus decomposition of $Z$.

Moreover, thanks to the domination constraint $M_t^\oplus \ge Z_t$ and identity (5.8), it comes that

$$u(N_t) \le M_t^\oplus \le u(N_t^*),$$

and thus, the running supremum processes of $M^\oplus$ and $Z$ are here indistinguishable: $M_t^{\oplus,*} = Z_t^* = u(N_t^*)$, for all $t \ge 0$.

These results can be summarized in the following proposition, where the filtration $(\mathcal{F}_t)$ is assumed to be quasi-left-continuous.

PROPOSITION 5.8. *Let $Z$ be a local supermartingale of the form $Z_t = u(N_t)$, where $u$ is an increasing concave function and $N$ a continuous local martingale such that $N_\infty = 0$. The Max-Plus decomposition of $Z$ is driven by the following processes:*

1. $Z_t = \mathbf{E}[L^*_{t,\infty}|\mathcal{F}_t]$ *and* $L^*_{0,t} = v(N^*_t)$, *where $v$ is an increasing function of the form $v(x) = u(x) - u'(x)x$ and $L_s = v(N^+_s)$.*
2. $M^\oplus_t = u(N^*_t) + (N_t - N^*_t)u'(N^*_t)$. *In particular, $M^\oplus$ and $Z$ have the same running supremum process:* $M^{\oplus,*}_t = Z^*_t = u(N^*_t)$.
3. *The optimal martingale $M^\oplus$ dominating the floor process $Z$, also satisfies the "stronger drawdown" constraint:* $M^\oplus_t \geq v \circ u^{-1}(M^{\oplus,*}_t)$.

*Previous revisited examples.*

• Note that the first case where $Z$ is a geometric Brownian motion with parameters $(-r, \sigma)$ can be included in the scope of the following example. In fact, $Z^\gamma$ (with $\gamma = 1 + \frac{2r}{\sigma^2}$) defines a martingale and so $Z_t = u(N_t)$, where $u(x) = x^{1/\gamma}$ is an increasing concave function and $N_t = Z^\gamma_t$. In particular, we can easily prove the result of Proposition 5.5 by simply using equation (5.8).

Moreover, we do not need Lemma 5.4 to prove that the exercise boundary $E^c(m) = m\mathbf{E}[\mathcal{Z}^*_{0,\infty}]$ of the perpetual American Call $C^{\mathrm{Am}}(Z,m)$ is nothing else $m\frac{\gamma-1}{\gamma}$. In fact, using the Azéma–Yor martingale (5.8), we get, for any time $t$,

$$\mathbf{E}[M^\oplus_t] = E\left[(N^*_t)^{1/\gamma} + \frac{1}{\gamma}(N^*_t)^{1/\gamma-1}(N_t - N^*_t)\right] = M^\oplus_0 = x.$$

Then taking the limit of the preceding expression as $t \to +\infty$ and since $N_\infty = 0$, we finally obtain the desired result.

More generally, let $N_t$ be a continuous exponential martingale of a stochastic integral such that $\int_0^\infty \sigma^2(N_t)\,dt = \infty$. Then the martingale associated with the Max-Plus decomposition of $Z_t = N_t^{1/\gamma}$ has the same form as the one in the geometric Brownian case.

• In the case of the considered Lévy processes, $Z_t = xe^{X_t}$, where $X$ has no positive jumps. So $Z^*_t$ is continuous and $N_t = Z^{\gamma \mathrm{L\acute{e}vy}}_t$ defines a martingale. Then the rest follows exactly as before.

• Let us now reconsider the case where $Z$ is a drifted Brownian motion with parameters $(-\mu, \sigma)$, where $\mu = r + \frac{\sigma^2}{2}$. It can be easily seen that $\exp(\gamma Z_t) = N_t$ defines a martingale and so $Z_t = u(N_t)$, where $u(x) = \frac{1}{\gamma}\log(x)$ defines an increasing concave function. Then in light of equation (5.8), the martingale $M^\oplus$ associated with the Max-Plus decomposition of $Z$ is of the form

$$M^\oplus_t = \frac{1}{\gamma}\log(N^*_t) + (N_t - N^*_t)\frac{1}{\gamma N^*_t} = Z^*_t + \frac{1}{\gamma}(\exp\gamma(Z_t - Z^*_t) - 1),$$

and we hence find the same formula as in Proposition 5.6. This property is not typical for the Brownian motion and remains valid for more general stochastic integrals.

REMARK 5.9. These results do not only hold under the assumptions $\zeta = \infty$ and $N_\infty = 0$. $M$ is still the martingale of the Max-Plus decomposition of $Z$, for any time horizon $\zeta$ such that $M_\zeta = v(N_\zeta^*)$, with $v$ a nondecreasing function.

5.4.2. *Max-Plus decomposition and American options.* Let us focus on the Snell envelope $Z_t(m)$ of $Z_t \vee m$, where $Z$ is of the form $Z_t = u(N_t)$. For this, we come back to the Azéma–Yor martingale and make it start from time $t$ instead of 0. Then, equation (5.8) becomes

$$M_s^t = u(N_{t,s}^*) + (N_s^t - N_{t,s}^*)u'(N_{t,s}^*) \qquad \forall s \geq t.$$

Since $M_t^t = \mathbf{E}[M_\infty^t | \mathcal{F}_t]$ and $N_\infty^t = 0$, we immediately obtain the Max-Plus decomposition of $Z$:

$$u(N_t) = \mathbf{E}[u(N_{t,\infty}^*) - N_{t,\infty}^* u'(N_{t,\infty}^*) | \mathcal{F}_t] = \mathbf{E}[v(N_{t,\infty}^*) | \mathcal{F}_t].$$

Now let us consider the Snell envelope $Z_t(m)$ of $Z_t \vee m$. In light of Section 4.3, $Z_t(m)$ is of the form

$$Z_t(m) = \mathbf{E}[\sup(v, m)(N_{t,\infty}^*) | \mathcal{F}_t].$$

Since $\phi(N_t) = \mathbf{E}[\phi(N_{t,\infty}^*) - N_{t,\infty}^* \phi'(N_{t,\infty}^*) | \mathcal{F}_t]$ for any increasing concave function $\phi$, if we find such a function $\phi$ satisfying $\sup(v, m) = \phi - x\phi'$, we immediately deduce that $Z_t(m) = \phi(N_t)$.

It can be easily seen that the function $\phi$ exists indeed and it is of the form $\phi(x) = m$ if $v(x) < m$ for any real $x$. In the contrary case, there exists some real $x_m^*$ satisfying $v(x_m^*) = m$ [by continuity of $\sup(v, m)$] and

$$\phi(x) = \begin{cases} u(x), & \text{if } x \geq x_m^*, \\ \dfrac{u(x_m^*) - m}{x_m^*}x + m, & \text{if } x < x_m^*. \end{cases}$$

Note that $x_m^*$ is well defined thanks to the nondecreasing property of $v$, and the function $\phi$ is nothing else but the concave envelope of $u \vee m$.

The following proposition precises the equivalence relation between the two functions $u$ and $v$.

PROPOSITION 5.10. *Let $Z$ be a local supermartingale of the form $Z_t = u(N_t)$, where $u$ and $N$ satisfy the same properties as in Proposition 5.8. Then $Z$ can be decomposed as follows:*

$$(5.9) \qquad Z_t = \mathbf{E}[v(N_{t,\infty}^*) | \mathcal{F}_t],$$

where $v$ is a nondecreasing function defined by $v(x) = u(x) - xu'(x)$.

Conversely, if we know that the supermartingale $Z$ admits the representation (5.9), we can deduce that $Z_t$ is of the form $Z_t = u(N_t)$, where the function $u$ solves the equation

$$u(x) - xu'(x) = v(x).$$

**6. Conclusion.** The paper suggests a new approach in martingale theory, which consists in looking for martingales under the form of a conditional expectation of some running supremum process. Such martingales are nothing else but an extension of the Doob–Meyer martingales in the Max-Plus algebra. The analysis of these martingales and their optimality property suggests a lot of potential applications to the theory of martingales and their maximum processes, via the Azéma–Yor martingale, in particular.

Moreover, our different point of view provides a unified framework for the solutions of many optimization problems related to the optimal stopping theory or the Bandit problem, and connects the notions of boundary and index process by means of the Max-Plus decomposition.

Hence, the Max-Plus decomposition of supermartingales and the uniqueness of the associated martingale turn out to be very useful in many optimization problems and have a lot of application fields, like American options and portfolio insurance in finance. We also think that our approach could be related to other works around the Max-Plus algebra (large deviations, . . . ).

## APPENDIX: PROOF OF PROPOSITION 4.2

We first observe that if $(a_i)_{i \in I}$ and $(b_i)_{i \in I}$ are two bounded families of real numbers,

$$\sup_{i \in I} a_i - \sup_{j \in I} b_j \le \sup_{i \in I} \left( a_i - \sup_{j \in I} b_j \right) \le \sup_{i \in I} (a_i - b_i),$$

and $|\sup_{i \in I} a_i - \sup_{j \in I} b_j| \le \sup_{i \in I} |a_i - b_i|$. Then since the function $m \mapsto x \vee m$ is 1-Lipschitzian,

$$|Z_S(m) - Z_S(m')| \le |m - m'|.$$

So we can define the regular random field $(Z_t(m); m \in \mathbb{Q})$ and make a continuous extension in $m$ from $\mathbb{Q}$ into $\mathbb{R}$.

The next step is to replace $m$ by an $\mathcal{F}_S$ random variable, bounded by below, and then to show that we still have

$$Z_T(\Lambda_S) = \operatorname*{ess\,sup}_{\tau \ge T} \mathbf{E}[Z_\tau \vee \Lambda_S | \mathcal{F}_T], \qquad T \in \mathcal{T}_{S,\zeta} \text{ a.s.}$$

First observe that the main properties of $(Z_S(m))_{S \in \mathcal{T}_{0,\zeta}}$ come from the fact that, for any stopping time $T \geq S$, the family $\{\mathbf{E}[Z_\tau \vee m | \mathcal{F}_S]; \tau \geq T\}$ is filtering nondecreasing. In particular,

$$(A.1) \qquad \mathbf{E}[Z_T(m) | \mathcal{F}_S] = \operatorname*{ess\,sup}_{\tau \in \mathcal{T}_{T,\zeta}} \mathbf{E}[Z_\tau \vee m | \mathcal{F}_S] \qquad \text{a.s.}$$

Now for any step random variable $\Lambda_S = \Sigma \mathbf{1}_{A_i} m_i$, with $(A_i)$ an $\mathcal{F}_S$-measurable partition of $\Omega$, let us define

$$\widetilde{Z}_T(\Lambda_S) = \operatorname*{ess\,sup}_{\tau \geq T} \mathbf{E}[Z_\tau \vee \Lambda_S | \mathcal{F}_T], \qquad T \in \mathcal{T}_{S,\zeta} \text{ a.s.}$$

In order to use property (A.1), we define a new stopping time $S_{A_i}$ that equals $S$ on $A_i$ and $\zeta$ on $A_i^c$. It comes that

$$\begin{aligned}
\mathbf{1}_{A_i} \widetilde{Z}_S(\Lambda_S) &= \mathbf{1}_{A_i} \mathbf{E}[\widetilde{Z}_{S_{A_i}}(\Lambda_S) | \mathcal{F}_S] \\
&= \mathbf{1}_{A_i} \operatorname*{ess\,sup}_{\tau \geq S_{A_i}} \{\mathbf{1}_{A_i} \mathbf{E}[Z_\tau \vee m_i | \mathcal{F}_S] + \mathbf{1}_{A_i^c} \mathbf{E}[Z_\zeta \vee \Lambda_S | \mathcal{F}_S]\} \\
&= \mathbf{1}_{A_i} \operatorname*{ess\,sup}_{\tau \geq S} \{\mathbf{E}[Z_\zeta \vee \Lambda_S | \mathcal{F}_S] \\
&\qquad\qquad + \mathbf{1}_{A_i} \mathbf{E}[(Z_\tau \vee m_i - Z_\zeta \vee m_i) | \mathcal{F}_S]\} \\
&= \mathbf{1}_{A_i} \mathbf{E}[Z_\zeta \vee m_i | \mathcal{F}_S] \\
&\qquad + \mathbf{1}_{A_i} \operatorname*{ess\,sup}_{\tau \geq S_{A_i}} \{\mathbf{E}[(Z_\tau \vee m_i - \mathbf{E}[Z_\zeta \vee m_i | \mathcal{F}_\tau]) | \mathcal{F}_S]\} \\
&= \mathbf{1}_{A_i} \operatorname*{ess\,sup}_{\tau \geq S_{A_i}} \mathbf{E}[Z_\tau \vee m_i | \mathcal{F}_S] = \mathbf{1}_{A_i} Z_S(m_i) \qquad \text{a.s.}
\end{aligned}$$

This means that $\widetilde{Z}_S(\Lambda_S) = Z_S(m_i)$ on $A_i$, and, therefore, $\widetilde{Z}_S(\Lambda_S) = \Sigma \mathbf{1}_{A_i} \times Z_S(m_i) = Z_S(\Lambda_S)$. The same argument may be applied at any stopping time $T \geq S$ to obtain that $Z_T(\Lambda_S) = \widetilde{Z}_T(\Lambda_S) = \operatorname{ess\,sup}_{\tau \geq T} \mathbf{E}[Z_\tau \vee \Lambda_S | \mathcal{F}_T]$, for any $T \in \mathcal{T}_{S,\zeta}$. Thanks to the Lipschitz property, this formula can be extended by continuity to any $\mathcal{F}_S$-measurable random variable bounded by below.

CMAP, Ecole Polytechnique
91128 Palaiseau Cedex
France
E-mail: elkaroui@cmapx.polytechnique.fr
          meziou@cmapx.polytechnique.fr